\documentclass[fleqn,usenatbib]{mnras}

\usepackage{newtxtext,newtxmath}

\usepackage[T1]{fontenc}

\DeclareRobustCommand{\VAN}[3]{#2}
\let\VANthebibliography\thebibliography
\def\thebibliography{\DeclareRobustCommand{\VAN}[3]{##3}\VANthebibliography}

\usepackage{graphicx}	
\usepackage{subcaption}
\usepackage{amsmath}	
\usepackage{anyfontsize}
\usepackage{adjustbox}
\usepackage{xcolor} 
\usepackage{enumitem}

\title[Optimizing Roman for chromatic PSF effects]{
Optimizing the \textit{Roman Space Telescope} High-Latitude Wide Area Survey for mitigating chromatic PSF effects on shear measurement}

\author[F. Berlfein et al.]{
Federico Berlfein,$^{1}$\thanks{E-mail: fberlfei@andrew.cmu.edu}
Rachel Mandelbaum,$^{1}$
Jiachuan Xu$^{2}$
and Tianqing Zhang$^{3}$
\\
$^{1}$ McWilliams Center for Cosmology and Astrophysics, Department of Physics, Carnegie Mellon University, Pittsburgh, PA 15213, USA \\
$^{2}$ Department of Physics, Northeastern University, Boston, MA 02115, USA\\
$^{3}$ Department of Physics and Astronomy and PITT PACC, University of Pittsburgh, Pittsburgh, PA 15260, USA
\\
}

\date{Accepted XXX. Received YYY; in original form ZZZ}

\pubyear{2015}

\begin{document}
\defcitealias{Berlfein_2025}{Paper~I}
\defcitealias{ROTAC_2025}{ROTAC Report}
\label{firstpage}
\pagerange{\pageref{firstpage}--\pageref{lastpage}}
\maketitle

\begin{abstract}
Chromatic point-spread-function (PSF) effects arise from differences between the spectral energy distributions (SEDs) of stars, used to model the PSF, and galaxies, used to measure shape distortions due to weak gravitational lensing, or shear. For the \textit{Roman Space Telescope}, these effects can bias shear measurement and cosmological inference, making them an important systematic effect for shear calibration. These biases depend sensitively on survey design choices, particularly filter coverage and the availability of color information. In this work, we investigate how different \textit{Roman} survey strategies affect the ability to mitigate chromatic PSF effects and whether residual biases in shear propagate into cosmological inference. Using realistic image simulations, we infer per-galaxy near-infrared SED slopes via radial basis function regression for four-, three-, two-, and single-band survey configurations. We quantify residual shear calibration biases under representative and non-representative training assumptions and propagate these biases into Markov Chain Monte Carlo analyses of cosmic shear and $3\times2$-point statistics. We find that three- and four-band strategies can reduce residual shear biases to $|m|\lesssim10^{-3}$, lowering the induced shifts in the lensing amplitude from $\Delta S_8 \sim 0.6\sigma$ (cosmic shear) and $\Delta S_8 \sim 0.7\sigma$ ($3\times2$-pt) in the uncorrected case to $\Delta S_8 \lesssim 0.07\sigma$.  
Single-band surveys remain intrinsically limited, with residual shear biases reaching or exceeding $|m|\sim 2\times 10^{-3}$ in some tomographic bins. Average, sample-wide corrections reduce but do not eliminate chromatic systematics, leaving residual biases of $\Delta S_8 \sim 0.1\sigma$. Overall, our results demonstrate that we can robustly correct for these effects in the recommended \textit{Roman} three-band medium tier, but may encounter residual biases in \textit{Roman}'s single-band wide tier. 
\end{abstract}





\section{Introduction}
\label{sec:intro}

Weak gravitational lensing (WL) is one of the most powerful probes of cosmology, providing a direct measurement of the large-scale distribution of matter through the coherent distortions, or shear, imprinted on the observed shapes of distant galaxies \citep{Hoekstra_2008, Bartelmann_2017}. 
By tracing both the expansion history and the growth of structure, WL enables stringent constraints on dark matter and dark energy. Recent surveys, including the Dark Energy Survey \citep[DES;][]{DES_2005}, Hyper Suprime-Cam Survey \citep[HSC;][]{Aihara_2017}, and Kilo-Degree Survey \cite[KiDS;][]{de_Jong_2012} have made precise measurements of cosmological weak lensing or cosmic shear, constraining the amplitude of matter fluctuations to the few percent level when assuming the $\Lambda$CDM cosmological model \citep[see][]{Asgari_2021,Amon_2022, Secco_2022, Dalal_2023, Li_HSC}. The next generation of surveys, such as the Vera C.\ Rubin Observatory Legacy Survey of Space and Time \citep[LSST;][]{LSST_2019}, \textit{Euclid} \citep{Euclid}, and the \textit{Nancy Grace Roman Space Telescope} High Latitude Imaging Survey \citep{Spergel_2015}, promise an order-of-magnitude improvement in statistical precision, enabling stronger tests of our cosmological model.

\textit{Roman} offers a particularly compelling opportunity for WL. \textit{Roman} combines 
a wide field of view with stable, diffraction-limited near-infrared (NIR) imaging providing a high angular resolution. This enables us to observe and distinguish between blended galaxies more effectively than with ground-based telescopes \citep{Troxel_2022}. This improvement simplifies many aspects of weak-lensing analyses, including galaxy detection, shape measurement, and perhaps most critically, photometric redshift inference. Reducing the number of unrecognized blends that would otherwise be treated as single objects can have a significant impact on accurate cosmological inference \citep{Nourbakhsh_2022}. 
In addition, photometric redshift estimation can be pushed to the high redshift range when combined with deep optical imaging from Rubin \citep{Eifler_2021}. The \textit{Roman} High-Latitude Wide Area Survey (HLWAS), as recommended by the 
\textit{Roman} Time Allocation Committee \citep[ROTAC;][hereafter \citetalias{ROTAC_2025}]{ROTAC_2025}, is expected to provide an unprecedented combination of depth, area, and wavelength coverage for weak lensing analyses. 

A key question is the survey strategy for WL with \textit{Roman}. Multiple observing scenarios (e.g., the filter choices, exposure time, tiling strategy, and joint coordination with Rubin) for the imaging portion of the HLWAS, known as the High-Latitude Imaging Survey (HLIS), have been active questions within the \textit{Roman} community \citep[e.g.,][]{Eifler_2021}. Scenarios under consideration have differed not only in the statistical power they provide, but also in the systematics we can calibrate for WL measurements. Survey parameters such as the choice of WL filters, the cadence of multi-band observations, and the availability of NIR color information directly affect photometric redshift performance, the calibration of shape systematics, and cosmological constraints \citep{ROTAC_2025}.

Because \textit{Roman} imaging is diffraction-limited, these choices also influence the wavelength-dependent properties of the point-spread function (PSF), creating a direct link between survey design and shear calibration. The accuracy of PSF modeling is a critical requirement for precision WL. Any mistimation of the PSF leads to biased shear measurements \citep{Mandelbaum_2018}, and for \textit{Roman}'s wavelength-dependent PSF the treatment of chromatic effects becomes especially important. In our first paper \citep[hereafter \citetalias{Berlfein_2025}]{Berlfein_2025}, 
we quantified the magnitude of shear biases arising from differences between the spectral energy distributions (SEDs) of stars, used to model the PSF, and galaxies, used to measure shear. We investigated a per-galaxy PSF-level mitigation approach and showed that, while chromatic biases are significant for \textit{Roman}, they can be reduced to within WL requirements when adequate information about the galaxy SED is available or can be inferred from photometric data.

In this follow-up work, we focus on the question of how survey strategy impacts the feasibility and performance of chromatic bias correction schemes. Because different strategies modify the depth, signal-to-noise ratio, and NIR color information available for each galaxy, they also change the accuracy with which each galaxy's effective PSF can be estimated. 
Strategies that provide multi-band NIR photometry enable more accurate SED inference, while strategies that rely on a single filter increase statistical constraining power due to increases in area, but limit systematic calibration. Similarly, joint \textit{Roman}–Rubin strategies produce combined optical+NIR color information, whose utility for chromatic mitigation can also depend on survey strategy. 

The two sources for survey configurations to explore are the \textit{Roman} Design Reference Mission \citep[DRM;][]{DRM_2021} and the \citetalias{ROTAC_2025}. The main differences between the DRM and the adopted strategy in the ROTAC report that affect WL are summarized in Table~\ref{tab:roman_survey_strategies}.  

\begin{table}
\centering
\caption{Summary of \textit{Roman} HLWAS strategies relevant for WL analyses. The DRM consists of a single wide area, 4-band imaging survey, while the \citetalias{ROTAC_2025} describes a 3-band medium tier optimized for calibration and a single-band wide tier optimized for area. We omit the HLWAS deep tier here since its area is insufficient for cosmological weak lensing. Depth here refers to the $5\sigma$ point-source  depth. 
}
\label{tab:roman_survey_strategies}
\begin{tabular}{lccc}
\hline
 & \textbf{DRM} & \multicolumn{2}{c}{\textbf{ROTAC Report}} \\
\cline{3-4}
 &  & \textbf{Medium Tier} & \textbf{Wide Tier} \\
\hline
Survey area [deg$^2$] & 2,000 & 2,400 & 2,700 \\
NIR filters & YJHF & YJH & H only \\
H-band depth (AB) & $26.7$ & $26.4$ & $26.2$ \\
Grism spectroscopy & Yes & Yes & No \\
\hline
\end{tabular}
\end{table}

The general motivation behind this change from the DRM was that the medium tier can focus on detailed calibration of WL systematics and photometric redshifts, while the wide tier's additional area is statistically beneficial for cosmology and has other general astrophysics utility.

Using the same image simulation suite and analysis methods as \citetalias{Berlfein_2025}, we evaluate the sensitivity of chromatic PSF estimation and shear calibration to key survey design choices. This work builds directly on feedback previously provided to the ROTAC and offers a detailed and quantitative justification for that input by assessing how alternative survey strategies affect chromatic correction performance and cosmological inference. Our analysis is structured around answering the following questions, with the first two focused on the medium tier, the next two on the wide tier, and the final question applicable to both tiers:

\begin{itemize}[leftmargin=*]
    \item \textbf{4 vs.\ 3 bands:} Does the removal of the F184 filter in the medium tier substantially affect our ability to correct for chromatic PSF effects?
    \item \textbf{Which 3 bands:} In the case of a 3-band medium tier, is there an important difference between YJH and JHF?
    \item \textbf{Single band correction:} How well can we correct for these effects for a single band wide tier survey, either by using LSST photometry or by applying an average correction derived using the medium tier?
    \item \textbf{Is two bands enough:} Is a 2-band wide tier (JH) enough to correct for these effects? 
    \item \textbf{Need for LSST:} Will we need LSST photometry to apply an accurate correction in the medium and/or wide tier?
\end{itemize}



The remainder of this paper is structured as follows. In Section~\ref{Background}, we review chromatic PSF effects in weak lensing, describe the \textit{Roman} survey strategies considered in this work, and summarize the simulation framework. Section~\ref{Mitigation} introduces the PSF-level mitigation formalism and the regression-based method used to infer galaxy SED information from photometry. In Section~\ref{Shear_Cal}, we quantify the resulting shear calibration performance across survey strategies and training-sample assumptions, and validate key trends using JWST/NIRSpec data. In Section~\ref{Cosmology} we propagate the residual shear biases into full cosmological analyses, assessing their impact on constraints from cosmic shear and joint $3\times2$-point statistics. We conclude in Section~\ref{Conclusion} with a summary of the main results and their implications for \textit{Roman} survey design and chromatic PSF mitigation.

\section{Background and Context}
\label{Background}



Weak lensing shear measurement requires an accurate characterization of the PSF \citep{Anderson_2000, Piotrowski_2013,Jarvis_2021, Liaudat_2023, Schutt_2025}. For \textit{Roman}, the PSF is strongly wavelength-dependent due to being diffraction-limited \citep{Yamamoto_2024}. Because the effective PSF relevant for a galaxy depends on its spectral energy distribution (SED), while the PSF is modeled using stars with different SEDs, star–galaxy SED differences introduce a chromatic bias in shear estimation \citep{Cypriano_2010, Eriksen_2018}. In this section we will recap the main results from \citetalias{Berlfein_2025} and the relevance of chromatic effects for \textit{Roman}, introduce the different survey scenarios we are considering, and the simulation suite used to test them.

\subsection{Previous Findings}
\label{sec:chromatic_psf_background}

In \citetalias{Berlfein_2025}, we quantified these biases for the four \textit{Roman} HLIS DRM weak lensing bands (Y106, J129, H158, F184) and for the much wider W146 filter using \textit{Roman}-like image simulations with realistic galaxy and stellar SEDs from the \textsc{cosmoDC2} \citep{Korytov_2019, Kovacs_2022} and \textsc{Diffsky} \citep{Hearin2020, OU_2025} extragalactic catalogs. 
More specifically, we measured how uncorrected chromatic PSF biases affect the observed shear relative to the true input shear in the simulations. The relationship between the observed and true shear is often expressed through a linear model \citep{Heymans_2006,Huterer_2006}:
\begin{equation}\label{eq:obs_shear}
\hat{\gamma} = (1 + m) \gamma + c,
\end{equation}
where $m$ and $c$ represent multiplicative and additive biases, respectively.

One key component of our input to the ROTAC was on the impact of chromatic PSF effects on the W146 filter. The W146 filter spans the full wavelength range covered by the four DRM WL filters and was originally proposed as an interesting option for WL due to its ability to reach significantly greater depth at fixed observing time \citep{Eifler_2021}. 
Using a single wide filter also significantly increases survey area at fixed observing time and for a given depth, 
improving the statistical cosmological power of WL. However, the broad wavelength coverage that increases depth greatly enhances the sensitivity of the PSF to galaxy SED variations. While a wide filter could substantially boost WL statistics, \citetalias{Berlfein_2025} shows that it dramatically increases the chromatic shear bias unless implausibly accurate chromatic corrections are available. 

We showed that star–galaxy SED differences produce  average multiplicative biases 
of order $\sim 0.25\%$ in the four DRM WL bands (Y106, J129, H158, F184), and substantially larger biases of order $\sim 2\%$ for the W146 filter. When divided into tomographic bins as will be done for cosmological WL analyses, the biases for the worst of the individual bins 
were as high as $\sim 0.6\%$ for the DRM bands and $\sim 4\%$ for the W146 filter. These values exceed both the \textit{Roman} SRD requirement ($|m| < 3\times10^{-4}$) and a relaxed requirement of $|m| < 10^{-3}$. These results were used to inform the ROTAC that a single-band wide tier survey using one of the DRM WL bands (in this case the H-band), could dramatically decrease biases due to chromatic effects when compared to the W146 filter. 


\begin{figure}
        \centering
        \includegraphics[width=1.0\linewidth]{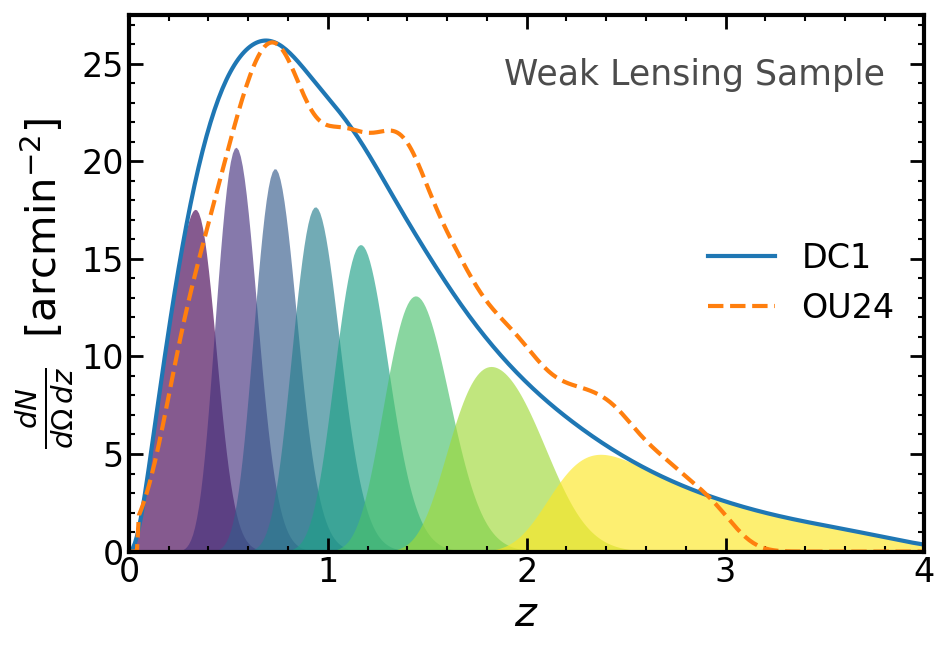}
        \caption{The redshift distribution for the weak lensing galaxy sample from the first \textit{Roman} Cosmology data challenge (blue) and our OU24 sample (orange) for comparison. There are a total of 8 tomographic bins, with the highest reaching $z = 4$ for DC1. We note that our OU24 sample reaches a maximum redshift of $z = 3.06$. For visualization purposes, the OU24 distribution is smoothed using a Gaussian kernel density estimator. 
        }
        \label{fig:n_z}
    \end{figure}

\subsection{Shear bias requirements}
\label{BiasRequirements}

In order to properly assess and contextualize the size of WL shear biases, we need mission- and science-motivated requirements. For \textit{Roman}, the only official requirement on shear multiplicative bias is set by the SRD to ensure that systematics remain sub-dominant to statistical uncertainties under worst-case assumptions. More specifically, the multiplicative shear requirement is derived from a conservative Fisher forecast of cosmic shear for a potential extended mission with 5 times the reference area at a reference depth. This conservative approach yielded a multiplicative bias requirement that can be considered stringent when compared to the requirements from other Stage IV surveys like LSST \citep[$|m| < 3\times10^{-3}$; ][]{LSST_2018} and Euclid \citep[$|m| < 2\times10^{-3}$; ][]{Massey_2012}. However, this was justified at the time as the goal was to provide guidance for early hardware production (e.g., the detectors) and it is generally the case that we want residual biases due to those to be a greatly subdominant part of the error budget. However,  fundamental features of the observatory that are irreducible in hardware and must be addressed using software motivate the introduction of the relaxed requirement in \citetalias{Berlfein_2025} as an alternative comparison for shear biases.

It is important to note that requirements on the multiplicative shear bias are generally defined for the \emph{total} contribution to $m$, combining multiple sources of systematic bias and uncertainty (e.g.\ PSF modeling, blending, detector effects, calibration residuals) in a root-mean-square sense. In this work, we isolate a single contribution to the overall shear bias budget, namely chromatic PSF effects. Directly comparing the magnitude of this individual contribution to the full requirement is therefore intentionally conservative and does not reflect the way shear-bias budgets are typically allocated across multiple effects. For this reason, we consider that using this relaxed requirement as a representative per-effect benchmark is justifiable. This value 
does not allocate the full systematic budget commonly adopted for other Stage~IV surveys to a single effect. Adopting this relaxed requirement allows us to then contextualize the chromatic PSF biases studied here with a reasonable benchmark for a Stage IV survey.


\begin{table}
    \centering
    \caption{Summary of survey scenarios considered in this work. WL cuts are based on  S/N $>18$ with combined J+H imaging (except for the Wide tier); resolution factor > 0.4; and ellipticity measurement error per component $\sigma_e < 0.2$  \citep[as defined in][]{Bernstein_2002}. 
    }
    \begin{tabular}{lccc}
        \hline
    \textbf{Scenario} & \textbf{Shorthand Label} & \textbf{Filters} &  \textbf{WL Cut} \\
        \hline
        DRM        & DRM & Y, J, H, F  &  H < 24.96 \\
        Medium Tier     & Medium & Y, J, H   & H < 24.60 \\
        Alternative Medium tier      & Medium-JHF & J, H, F        &   H < 24.60 \\
        Wide tier     &  Wide & H       &  H < 24.10\\
         Alternative Wide tier     &  Wide-JH & J,H      &  H < 24.50 \\
        \hline
    \end{tabular}
    \label{tab:survey_scenarios}
\end{table}

\subsection{Survey Scenarios}
\label{SurveyScenarios}

The final recommendations for the observing strategy for HLWAS are part of the latest \citetalias{ROTAC_2025}. Different strategies were considered, varying in filters, exposure time, and area, among others changes. All of these design choices have direct consequences for photometric redshift inference and weak lensing calibration. In this work, we consider a set of survey scenarios, ranging from multi-band NIR imaging to a single-band H-only strategy, with the goal of understanding how we can optimize the \textit{Roman} filter selection to correct for chromatic PSF effects to within the requirements.

Table.~\ref{tab:survey_scenarios} summarizes the differences between these strategies, mainly in filter choice and approximate magnitude cuts for WL selection. Different survey strategies affect the available SED information for galaxies and therefore the accuracy of chromatic PSF corrections. Strategies with more NIR color information (e.g. DRM or medium tier survey) allow more precise estimation of the effective within-band SED slope of each galaxy, 
enabling accurate corrections to the PSF. Conversely, single-band approaches (e.g. wide tier) lack direct NIR color constraints, assuming that either (a) the addition of optical photometry from Rubin LSST will be predictive of the NIR SED, or (b) we can apply an average calibration from the deep/medium tiers to the wide tier. As we show in later sections, these differences in strategy propagate directly into the achievable shear calibration accuracy and cosmological inference.

For all survey scenarios, we divide each sample into 8 tomographic source bins as shown in Fig.~\ref{fig:n_z}. These bins are defined in  \textit{Roman}'s HLIS Data Challenge 1 (DC1) and described in its companion paper (J. Xu et al.\ in preparation) and in \cite{Cao_2026}. More details on DC1 can be found in Sec.~\ref{Cosmology}. For comparison, we also show our OpenUniverse2024\footnote{\url{https://irsa.ipac.caltech.edu/data/theory/openuniverse2024/overview.html}} \citep[][hereafter OU24]{OU_2025} medium-tier selected WL sample. We notice that our OU24 sample only goes up to $z = 3.06$ and does not exactly match the DC1 distribution, which is expected.

\subsection{Use of LSST Photometry}
\label{sec:lsst_photometry}


Throughout this work we consider both \textit{Roman}-only and LSST$+$\textit{Roman} configurations when inferring galaxy SED information for chromatic PSF correction, where we assume LSST Year 4  depth in the combined case. 
It is important to emphasize, however, that the use of LSST photometry for chromatic mitigation enters at a much earlier stage in the image processing pipeline when compared to its more commonly discussed role in photometric redshift estimation. Chromatic PSF correction must be applied prior to shape measurement, since it directly affects the effective point-spread function used to measure galaxy shapes. By contrast, external photometry is typically incorporated at a much later stage, once galaxy catalogs have been constructed and multi-band measurements are combined to estimate photometric redshifts. As a result, using LSST data to inform chromatic PSF corrections would require integrating external optical photometry (or products derived from it) into the \textit{Roman} image-processing pipeline much earlier than would otherwise be necessary.

This distinction makes the LSST$+$\textit{Roman} chromatic correction scenario genuinely non-trivial. It would require reliable cross-survey matching and consistent photometric calibration prior to shape measurement. For example, the substantial difference in imaging resolution between LSST and \textit{Roman} complicates cross-survey star selection. Stars that are cleanly resolved in \textit{Roman} may be blended in LSST, leading to errors in the photometry or requiring exclusion of potential PSF stars from the sample that we would otherwise want to keep. While such an approach may ultimately be feasible, it carries additional operational and systematic complexity relative to \textit{Roman}-only corrections based on NIR photometry. For this reason, we explicitly evaluate both \textit{Roman}-only and LSST$+$\textit{Roman} cases in this work. This allows us to assess the extent to which chromatic PSF effects can be mitigated using \textit{Roman} data alone, and to quantify the potential gains associated with incorporating LSST photometry at this early stage of the weak-lensing pipeline.

\subsection{Simulations and shear estimation}
\label{Simulations}

We build upon the simulation framework\footnote{\url{https://github.com/Roman-HLIS-Cosmology-PIT/RomanChromaticPSF}} developed in \citetalias{Berlfein_2025}. 
The galaxy and stellar populations are drawn from the same \textsc{Diffsky} extragalactic and stellar catalogs used for OU24. As described in \citetalias{Berlfein_2025}, we apply catalog-level noise to the true \textit{Roman} and LSST magnitudes using the photometric error model in \texttt{PhotoErr v1.3} \citep{crenshaw2024}. This is done for a deep-like and wide-like survey (wide-like here includes the DRM, the medium tier, or the wide tier), representing the training and test sets respectively. 
The training set consists of 40,000 galaxies and the test set has 10,000 galaxies (for which we generate images) that pass the WL magnitude cuts. We also highlight that the training and test set have different noise levels, as they represent the deep tier and either the wide or medium tier, respectively. 

We generate the image simulations using \textsc{GalSim v2.6} \citep{Galsim}, a widely used Python framework for astronomical image simulation. 
In particular, we rely on the \textit{Roman} module within \texttt{GalSim}, which incorporates detailed instrument characteristics and PSF models specific to the \textit{Roman} observatory. Because the image-level simulations remain unchanged from Paper~I, no new images are generated for this work. Instead, for each survey strategy we recompute catalog-level magnitude errors and derive new signal-to-noise (SNR) estimates appropriate for the altered depths. We then apply strategy-specific WL selection cuts, selecting only those simulated galaxies that would pass the SNR threshold of each scenario. All image-level shear measurements therefore rely on a consistent set of \emph{noiseless} image simulations. This allows us to use the same images for each survey scenario, while the weak lensing samples themselves differ across strategies. Finally, we restrict ourselves to shape measurement in the H-band only, as this is the only filter common to all survey strategies and scenarios. 

We measure galaxy shapes with \textsc{AnaCal} \citep{Li_2023, Li_2024b}, a framework for analytical shear estimation using Fourier Power Function Shapelets \citep{Li_2018, Li_2022b}. \textsc{AnaCal} includes modules for handling realistic observational complications such as galaxy detection, sample selection, and noisy observations. In this work, however, the simulated images are noiseless and galaxies are force-detected to eliminate detection-related biases and isolate chromatic effects. Consequently, we make use only of \textsc{AnaCal}’s shape-measurement component, applied to isolated galaxies in noiseless images

For further details on the mock catalogs, photometric error model, image simulations, and shear estimation via \textsc{AnaCal}, we refer the reader to \citetalias{Berlfein_2025}.

\begin{figure}
        \centering
        \includegraphics[width=1\linewidth]{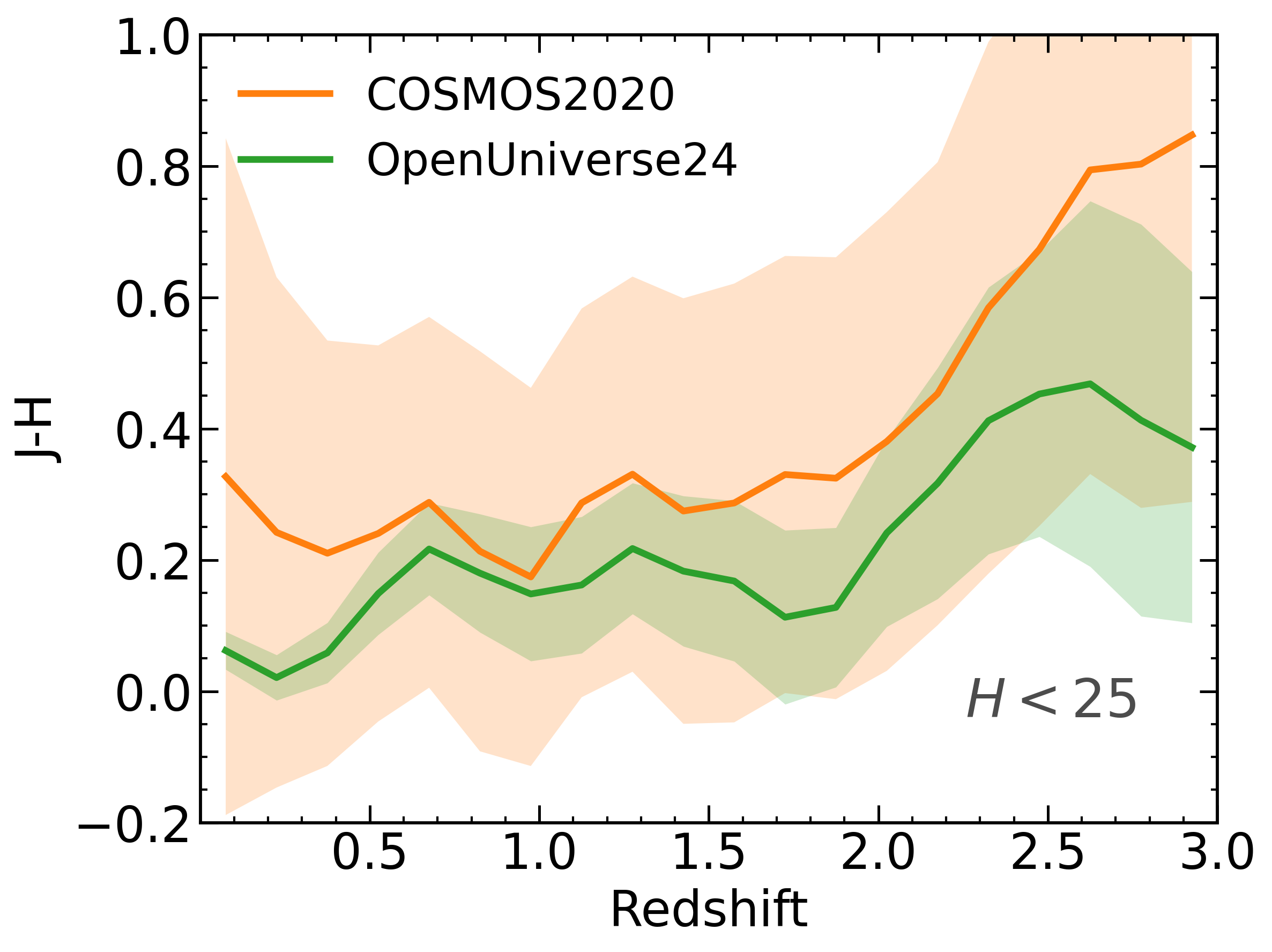}
        \caption{
        Mean $J-H$ color as a function of redshift for galaxies with $H<25$, comparing the OpenUniverse2024 (green) catalog with added photometric noise to the COSMOS2020 Farmer catalog (orange). Solid lines denote the mean $J-H$ color when binned over redshift, while the shaded regions indicate the $1\sigma$ scatter. We can clearly see that at most redshifts, the OU24 NIR color-redshift relation is much tighter than that observed in real data. We confirm that the spread of the COSMOS2020 $J-H$ color is not driven by outliers, but rather it is due to features of the real data that are not present in simulations. 
}

        \label{fig:color_vs_z}
    \end{figure}

\subsection{Known Issues}
\label{subsec:knownissues}
It is worth highlighting two known limitations in the extragalactic catalogs used in Paper~I: (1) repeated SEDs for galaxies,  
and (2) narrower-than-realistic NIR color variance \citep{OU_2025}. This can be seen in Fig.~\ref{fig:color_vs_z}, where we compare the redshift vs.\ $J-H$ color of the OU24 galaxies and the COSMOS 2020 galaxies from the Farmer catalog \citep{COSMOS2020} 
We see much tighter colors in OU24 than in COSMOS, indicating a lack of variance in the NIR SEDs. This is an important detail when interpreting the results from the simulations in this work, where the variation in SEDs is not as wide as what we will have in real data. This means that the biases we measure using OU24 are likely smaller than those we expect from real data. We try to address these limitations in Sec.~\ref{sec:rbf_regression} and contextualize our results with real data in Sec.~\ref{sec:jwst_validation}.

\section{Mitigation Method}\label{Mitigation}

In this section we will recap the mitigation method introduced in Paper I to correct for chromatic effects at the PSF level, as well as our chosen method to infer the NIR SED information from photometric data.

\subsection{Mitigation Method}

In Paper~I, chromatic modeling was framed through the definition of the per-galaxy effective PSF:
\begin{equation}
    \mathrm{PSF}_{\mathrm{eff}, o}(x,y)
    =
    \frac{\int \mathrm{d}\lambda\, \mathrm{PSF}(x,y,\lambda)\,F(\lambda)\,\mathrm{SED}_o(\lambda)}
    {\int \mathrm{d}\lambda\, F(\lambda)\,\mathrm{SED}_o(\lambda)},
    \label{eq:effective_psf}
\end{equation}
where $F(\lambda)$ is the filter throughput and $\mathrm{SED}_o(\lambda)$ is the normalized SED of object $o$. Since galaxies and stars have different SEDs, their effective PSFs differ:
\begin{equation}
    \Delta \mathrm{PSF}_{\rm eff}(x,y)
    = 
    \mathrm{PSF}_{\rm eff,\star}(x,y)
    -
    \mathrm{PSF}_{\rm eff, g}(x,y),
\end{equation}
and these differences propagate into shear measurement biases if not accounted for.

To enable chromatic PSF corrections, Paper~I introduced a PSF-level mitigation formalism based on expanding the SED difference between stars and galaxies in a Taylor series around the filter effective wavelength $\lambda_0$:
\begin{equation}
    S_o(\lambda) 
    =
    \sum_{n=0}^\infty 
    S_o^{(n)} 
    \frac{(\lambda - \lambda_0)^n}{n!},
\end{equation}
leading to a decomposition of the chromatic PSF error into a linear combination of basis images:
\begin{equation}
    \Delta \mathrm{PSF}_{\mathrm{eff}}(x,y)
    =
    \sum_{n}
    \Delta S_n \, B_n(x,y),
    \label{eq:basis_expansion}
\end{equation}
where the coefficients $\Delta S_n$ capture the differences in the SED derivatives of stars and galaxies, and the basis images:
\begin{equation}
    B_n(x,y) 
    =
    \int \mathrm{d}\lambda\, 
    \mathrm{PSF}(x,y,\lambda)\,F(\lambda)\,(\lambda-\lambda_0)^n
\end{equation}
depend only on the PSF model and filter throughput, and not on the SED. 

This decomposition is central for the present work because it makes explicit where the SED information enters the correction: the galaxy’s contribution appears only through the coefficients $\Delta S_n$, which must be inferred from photometric data. It is worth highlighting that we primarily care about the SED slope {\em at the effective wavelength of the filter}, which is what we attempt to infer by using broadband colors (e.g. $J-H$). In \citetalias{Berlfein_2025}, we demonstrated that (i) the chromatic bias is dominated by the first-order term ($n=1$), and (ii) estimating $\Delta S_1$ from NIR colors allows the corrected PSF to recover shear within shear calibration requirements for  DRM-like survey configuration.

The present analysis builds directly on this formalism. However, unlike Paper~I, which assumed DRM-like 4-band photometry, we now evaluate how different survey strategies modify the available SED information and therefore the accuracy with which each galaxy's $\Delta S_n$ can be inferred, assessing the impact on the shear calibration bias $m$ and propagating that into cosmological constraints.

\subsection{RBF Ridge Regression}
\label{sec:rbf_regression}


To infer the SED-dependent coefficients $\Delta S_n$ from photometric data, we use a regression model based on radial basis functions \citep[RBF's;][]{Broomhead_1988}, as implemented in the \textsc{scikit-learn} \citep{scikit-learn} module \textsc{RBFSampler}. 
In \citetalias{Berlfein_2025} we used a self-organizing map (SOM) to infer SED-dependent corrections from galaxy colors. However, we found that this approach produced overly optimistic results when applied to the OU24 catalog, whose narrow NIR color distributions allowed the SOM to effectively memorize the training sample. To address this limitation, we use an RBF model which enforces smoothness in the mapping from the features (colors) to the target (SED slope) \citep{Rasmussen_2006}. We found this approach to be less susceptible to overfitting in narrow color spaces. 

An RBF is a function whose value depends only on the distance between two points, typically through a norm $\|\mathbf{x}-\mathbf{x}'\|$. The Gaussian RBF, the most commonly used form, is defined as
\begin{equation}
    k(\mathbf{x},\mathbf{x}')
    = \exp\!\left(
        -{\gamma\|\mathbf{x}-\mathbf{x}'\|^2}
    \right),
    \label{eq:rbf_kernel}
\end{equation}
where $\gamma$ controls the scale over which the function varies. Kernels of this form enable non-linear regression using simple linear models in a higher-dimensional feature space. In this feature space, smooth non-linear relationships between colors and SED properties can be represented as linear combinations of basis functions centered on the training samples.

Direct kernel regression, however, requires constructing and manipulating an $N\times N$ kernel matrix for $N$ training samples, which is computationally expensive for the large datasets \citep{Rasmussen_2006}. For this reason, we use the random Fourier feature approximation of the RBF kernel introduced by \cite{Rahimi_2007} and implemented in \textsc{scikit-learn}. This method relies on the fact that a shift-invariant kernel such as Eq.~\eqref{eq:rbf_kernel} can be expressed as the Fourier transform of a probability distribution. By sampling from this distribution, one can construct an explicit, finite-dimensional feature map that approximates the kernel:
\begin{equation}
    z_j(\mathbf{c})
    = \sqrt{\frac{2}{D}} \,
      \cos\!\left(\boldsymbol{\omega}_j^\top \mathbf{c} + b_j \right),
\end{equation}
where $\mathbf{c}$ denotes the input feature vector (e.g., its observed colors), the frequencies $\boldsymbol{\omega}_j$ are drawn from a normal distribution with variance $\gamma^{2}$, the phases $b_j$ are drawn uniformly from $[0,2\pi)$, and $D$ is the dimensionality of the feature map. The inner product between two such feature vectors approximates the RBF kernel:
\begin{equation}
    \mathbf{z}(\mathbf{c})^\top \mathbf{z}(\mathbf{c}')
    \approx
    k(\mathbf{c},\mathbf{c}').
\end{equation}

With this mapping, non-linear regression becomes a standard linear regression problem in the transformed feature space. We use ridge regression on the RBF features to predict the target SED slope:
\begin{equation}
    \hat{y}
    = \boldsymbol{\beta}^\top \mathbf{z}(\mathbf{c}) +\alpha \sum_i^N \boldsymbol{\beta}_i ,
\end{equation}
where $\boldsymbol{\beta}$ is the vector of regression coefficients and $\alpha$ is a regularization penalty on $\boldsymbol{\beta}$ that controls model complexity and prevents overfitting. 

This method is well suited to the OU24 extragalactic catalog, which contains repeated or nearly duplicated SEDs in the mocks. Tree-based regression models, including random forests and gradient-boosted trees, tend to ``memorize'' such duplicated samples and learn sharp, non-physical decision boundaries in feature space \citep{Breiman_2001, Friedman_2001}. In contrast, the RBF regression enforces smoothness: the Gaussian kernel imposes a prior that function values vary smoothly with the features \citep{Rasmussen_2006}.This makes the method more robust to duplicated entries and less prone to memorization issues.

\begin{figure*}
        \centering
        \includegraphics[width=0.99\linewidth]{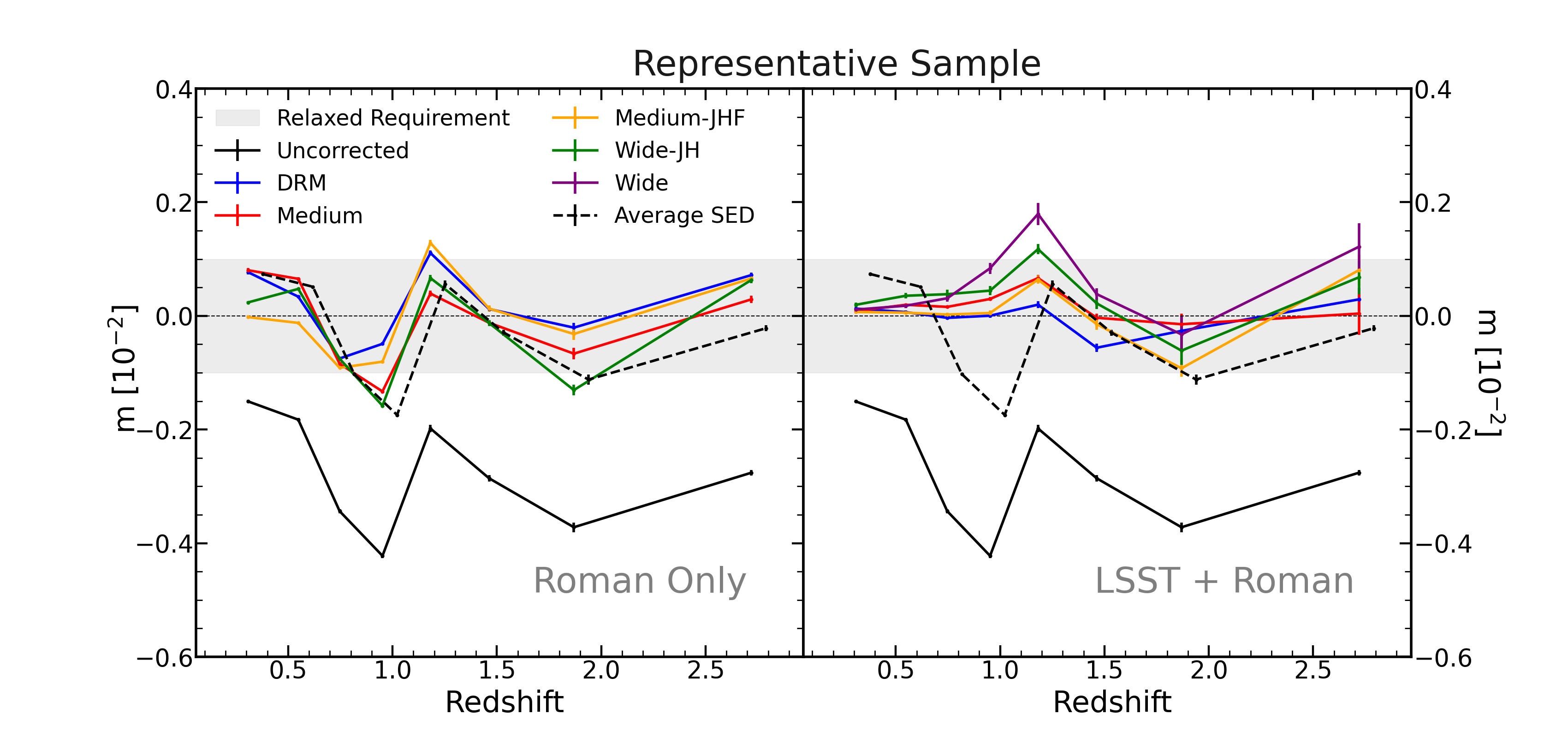}
        \caption{
        Shear multiplicative bias $m$ as a function of redshift for the 8 \textit{Roman} tomographic bins, measured from $H$-band shape measurements. Black solid lines show the uncorrected bias, while the black dashed lines correspond to applying a single true average correction to all galaxies. Colored lines indicate the residual bias after applying survey-specific corrections inferred using the RBF method with a representative training sample to predict the $H$-band SED slope from available photometric information. The left panel is for an analysis of \textit{Roman} photometry alone, while the right panel uses LSST and \textit{Roman} photometry. For the Wide tier, which has only a single \textit{Roman} NIR band, an SED-based correction cannot be inferred on a per-object basis from \textit{Roman} data alone. Instead, we can apply an average correction calibrated from the Medium tier, or use complementary LSST photometry to infer the SED-based correction for every galaxy. For this reason, the per-galaxy Wide-tier result is shown only in the right panel. 
        Shaded gray bands indicate the relaxed requirement, described in Sec.~\ref{BiasRequirements}, for comparison. 
        We can see that most strategies fall within or close to the relaxed requirement when including LSST photometry, with the exception of the wide tier. When only \textit{Roman} colors are used, the correction can fall outside of requirements for some redshift bins, but without any systematic trend towards positive or negative biases, with the exception of Wide-JH.
}

        \label{fig:mbias_repsample}
    \end{figure*}

In this work we fix the number of random Fourier components to $D = 1000$, as we found our results do not vary if we increase this further. Much smaller values ($D < 100$) can lead to noticeable underfitting. We optimize the ridge-regularization strength ($\alpha$) and kernel scale ($\gamma$) using a 90/10 train--test split drawn from a sample of 40{,}000 galaxies, selecting the hyperparameters that minimize the mean absolute error (MAE) on the held-out test set. We allow both parameters to vary between 0.00001 and 1.

To assess the robustness of the model to non-representative training data, we repeat the training using a bright subsample of galaxies with magnitudes $i < 24$, which represents a realistic scenario in which the the galaxies for which we have accurate SED information is brighter than our WL sample. This test evaluates whether the regression method can generalize when the training set is not fully representative of the test set.

\section{Shear Calibration Results}\label{Shear_Cal}

In this section, we present results for how well we are able to produce well-calibrated weak lensing shear estimates for two cases: a representative training sample, in which the training SEDs closely match the WL galaxy population, and a bright training sample, which reflects a more realistic scenario in which accurate SED information is primarily available for brighter or lower-redshift galaxies. For each survey strategy summarized in Table~\ref{tab:survey_scenarios}, we examine the residual multiplicative shear bias after applying the RBF-based chromatic correction based on the available color information. We show results both without and with LSST photometry, as we try to understand whether optical photometry is needed to infer the NIR SED slope accurately. We quantify performance of the correction method by measuring the residual shear multiplicative bias after applying a per-galaxy PSF correction based on the inferred SED slope. We present the multiplicative bias for each of the 8 tomographic bins in \textit{Roman} for each survey configuration. 
We compare the biases with the  relaxed requirement  $|m| < 1\times10^{-3}$ (Sec.~\ref{BiasRequirements}). 

\subsection{Representative Training Sample}
Figure~\ref{fig:mbias_repsample} shows the residual multiplicative shear biases in tomographic redshift bins for all survey strategies, assuming a fully representative training sample. The left panel is for \textit{Roman} photometry, while the right panel shows results using LSST and \textit{Roman} photometry. In both panels we also see that the size of the uncorrected effect usually varies with redshift between $m \sim 0.2\% - 0.4\%$, which clearly exceeds our shear calibration requirements. 
We also show the quality of a 
correction where the true average galaxy SED is known for the whole sample across all tomographic bins, for which biases can be mostly mitigated within requirements, with the exception of two tomographic bins. This shows that if we can accurately estimate the average SED slope of our WL sample, through the average color for example, the chromatic correction may be good enough in a subset of cases.  


\subsubsection{DRM and Medium tiers}
For all 3+ band strategies: DRM (YJHF), medium (YJH), and the medium alternative (JHF), our method achieves residual shear biases that lie within or very near the relaxed \textit{Roman} requirement with and without LSST photometry for all tomographic bins. 
In addition, we see no systematic trend towards positive or negative biases with either configuration. In addition, the difference between YJH and JHF is marginal: both provide enough NIR color information to accurately infer the $H$-band SED slope using the RBF regression, leading to similarly efficient chromatic PSF corrections. The DRM configuration performs slightly better in the case of no LSST photometry, but the differences between DRM and medium configurations are also minimal. 

LSST photometry provides a small improvement across survey configurations. In the LSST$+$\textit{Roman} panel, most strategies show slightly tighter biases than in the \textit{Roman}-only panel. 
The \textit{Roman}-only scenario leads to visibly larger scatter and to points for some tomographic bins falling outside the relaxed requirements. Nevertheless, the overall conclusion is that LSST photometry can be used to improve the correction, but using NIR photometry from \textit{Roman} alone is most likely sufficient.

\subsubsection{Wide tier}
The ROTAC-recommended wide-tier (H-only) shows the largest residual biases among the survey strategies. Since only a single NIR band is available, LSST photometry is required to enable a per-galaxy chromatic correction. The residual biases exceed the relaxed requirement in two tomographic bins. This highlights a key limitation of single-band weak-lensing surveys: without direct NIR color constraints, the NIR SED slope is difficult to infer. Comparing the single-band strategy to the two-band alternative (JH) illustrates this effect clearly. When LSST photometry is available, the two-band configuration more accurately constrains the NIR SED slope, and the resulting multiplicative biases largely remain within the relaxed requirement, with only one redshift bin slightly exceeding it. In contrast, for the single-band strategy without LSST photometry, the correction becomes more erratic, and two tomographic bins exceed the relaxed requirement. 

While the inclusion of LSST photometry can significantly improve performance in the wide tier, it is important to note that this improvement comes at the cost of increased pipeline complexity. As discussed in Section~\ref{sec:lsst_photometry}, using LSST data for chromatic PSF correction requires incorporating external photometry at an unusually early stage of the \textit{Roman} image-processing pipeline, making this scenario operationally and systematically more challenging than \textit{Roman}-only corrections.

Finally, it is important to emphasize that the representative-sample assumption is optimistic and unlikely to be realized in practice. In real data, galaxies in the WL sample are fainter, redder, and more diverse than those for which reliable spectroscopic or SED information is typically available.  Thus, the results in Figure~\ref{fig:mbias_repsample} should be interpreted as demonstrating performance for each survey strategy under ideal training conditions and highlight that having no NIR colors may impact our ability to accurately predict the NIR slope to the levels of shear calibration we care about. We use the average correction to assess the level of performance achievable when transferring calibration information from the medium tier to the wide tier. Because the wide tier lacks NIR color information, a practical approach is to derive an average correction from the medium tier and apply it to the wide data. This approximation may be sufficient in practice if we find residual biases across tomographic bins to remain within requirements.

\subsection{Bright Training Sample}
\begin{figure*}
        \centering
        \includegraphics[width=0.99\linewidth]{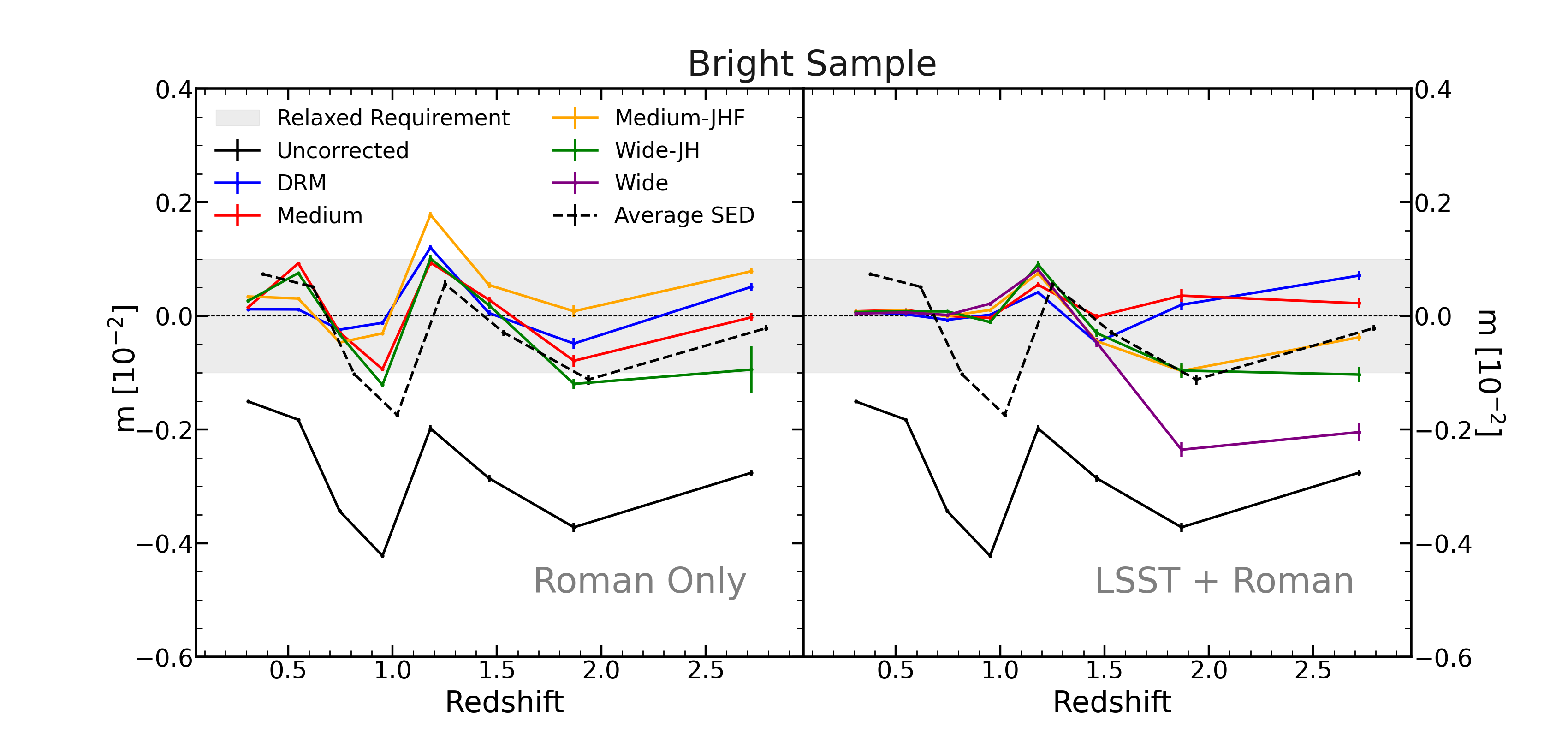}
        \caption{Similarly to Fig.~\ref{fig:mbias_repsample}, but with a bright training sample ($ i < 24$) for the RBF correction. We see a clear systematic residual bias for the wide tier at higher redshifts, where the training sample has less coverage. Other configurations with 2+ NIR bands can stay within or close to the relaxed requirement boundary in the case where both \textit{Roman} and LSST photometry is used for the correction. In the case of \textit{Roman}-only information, all 3+ band strategies perform similarly, with only a single tomographic bin in the Medium-JHF scenario exceeding the requirements. For the Wide-JH scenario, biases are found outside or at the requirement in three tomographic bins.   }
        \label{fig:mbias_brightsample}
    \end{figure*}

Figure~\ref{fig:mbias_brightsample} summarizes the residual shear biases when the regression model is trained on a bright subsample ($i<24$), representing a more realistic scenario in which the available spectroscopic training data are brighter than the full WL source sample. The cut is applied in the $i$ band, rather than in the $H$ band, because the majority of the existing spectroscopic surveys select targets based on the optical colors. Unlike the representative sample, the bright training set lacks high-redshift and intrinsically faint galaxies, resulting in a mismatch between the training and test distributions. After applying this cut, roughly 75\% of galaxies in our training set are discarded.

\subsubsection{DRM and Medium tiers}
For multiband NIR surveys (DRM, Medium, Medium-JHF), the RBF regression remains robust. Across redshift bins, the residual biases remain below the relaxed \textit{Roman} requirement, with the exception of a single bin in the Medium-JHF configuration. The degradation is most apparent at high redshift, where the training set provides limited coverage. The results and conclusions do not vary much when we exclude the LSST photometry, highlighting that this type of correction can most likely be done with \textit{Roman} photometry only. Once again, we do not see notable differences between the DRM, Medium, and Medium-JHF scenarios, showing that we can correct for chromatic effects within \textit{Roman} requirements with any 3+ band survey.

\subsubsection{Wide tier}
For the wide-tier H-only survey, residual shear biases reach $\sim 2\times10^{-3}$ at higher redshift bins, exceeding the relaxed requirement. This shows that the wide tier remains the most systematically challenging configuration, especially when the training data is not representative. Adding one band (Wide-JH scenario) can reduce the residual biases by a factor of two, and remain at the edge of the relaxed requirement in the LSST$+$\textit{Roman} and \textit{Roman}-only cases. Therefore, in the case where we find that these biases exceed future systematic requirements on the wide tier, adding J-band imaging could improve our ability to correct for chromatic effects. It is important to note that even though we see tight constraints at lower redshifts, this may not be the case with real data. If at lower redshifts our training data is not fully representative, biases for the wide tier might be larger. These  results simply highlight that in the event of a non-representative training sample, the lack of NIR colors in the wide tier could present a challenge for mitigation chromatic effects, while this effect seems to be smaller in the case where NIR colors are available.

\begin{figure*}
    \centering

    
    \includegraphics[width=0.45\linewidth]{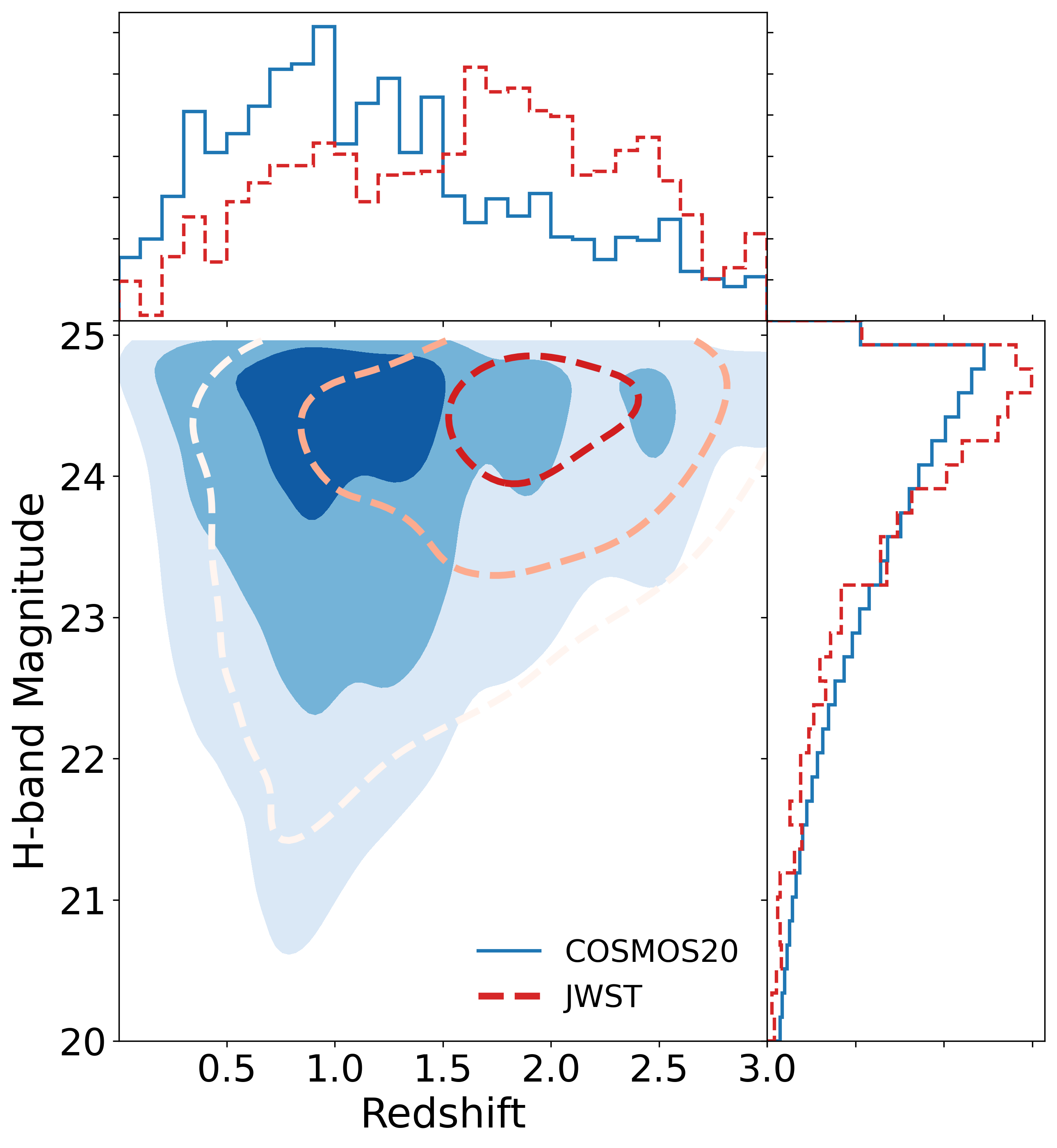}
    \includegraphics[width=0.45\textwidth]{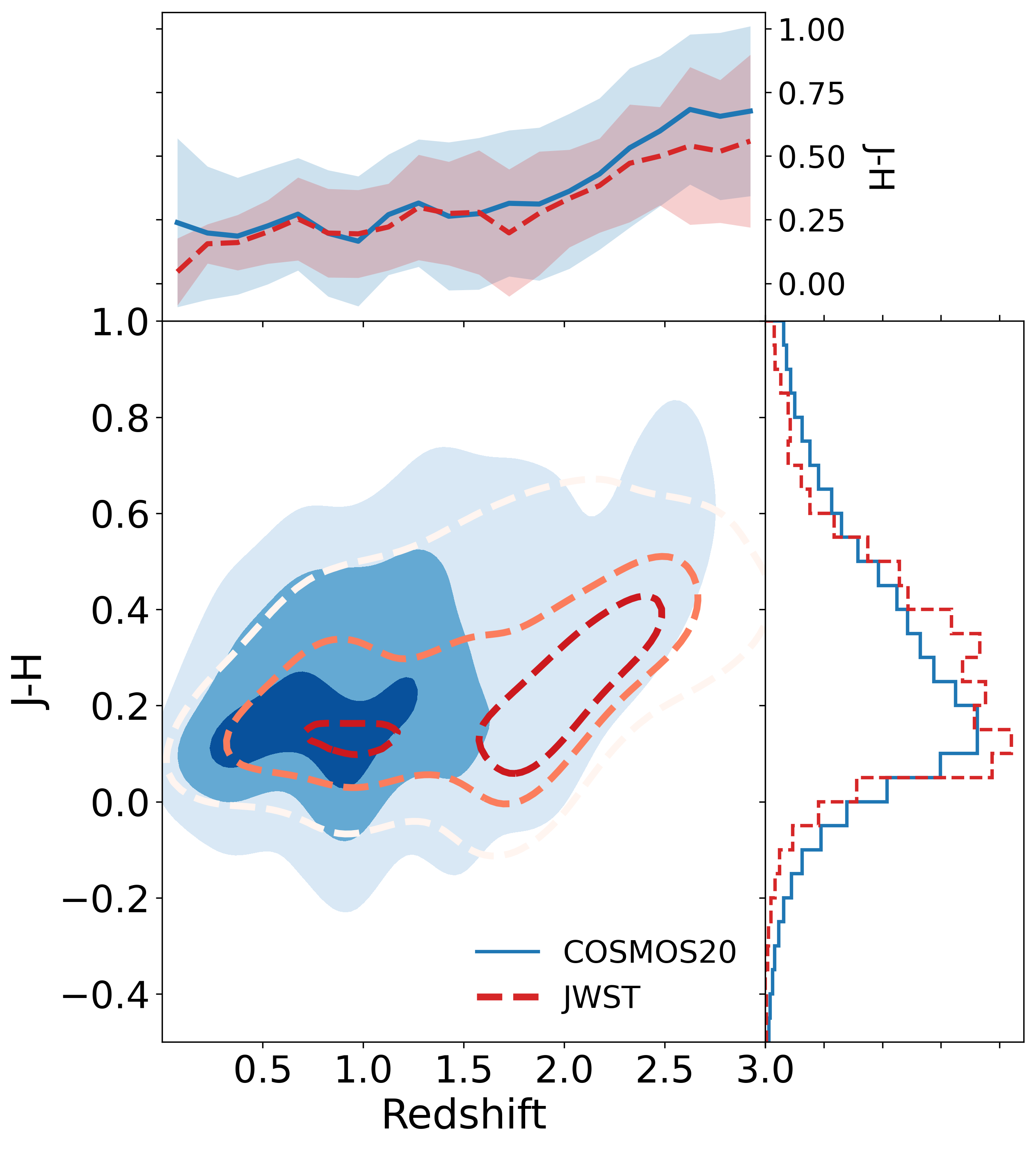}
    \caption{
    \textbf{Left}: Contour plot of the redshift vs.\ $H$-band magnitude for COSMOS20 (filled blue) and JWST NIRSpec (dashed red) galaxies. The adjacent 1D histograms show the respective distributions.  \textbf{Right}: Contour plot of the redshift vs.\ $J-H$ color for the same galaxies. The upper panel shows the $J-H$ color and its $1\sigma$ dispersion as a function of redshift. The JWST galaxies have a higher average redshift, yet follow similar color-redshift evolution and occupy a similar region of color space. All 2-D contours show 16th, 50th, and 84th percentiles of the data distribution.
}

    \label{fig:jwst_cosmos_comp}

\end{figure*}

\subsection{Comparisons with JWST NIRSpec}
\label{sec:jwst_validation}

Given the known limitations of the OpenUniverse24 simulations (Sec.~\ref{subsec:knownissues}), we wish to compare the color--SED relationships inferred from the simulations to those of real galaxies.  For this purpose, we compare our results with the simulations against those for a spectroscopic sample drawn from \textit{JWST} NIRSpec \citep{NIRSpec_paper}. We use the third version of a public NIRSpec dataset\footnote{\url{https://dawn-cph.github.io/dja/spectroscopy/nirspec/}} \citep{Brammer_2022,Heintz_2024, de_Graaff_2025}, composed of observations from a multitude of independent programs: 
JADES, CEERS, UNCOVER, RUBIES, CAPERS, GOODS-N, PEARLS, and several cluster-lensing campaigns (e.g., Abell 2744, MACS\,J0647). These programs span a wide range of scientific goals, target-selection criteria and depths. As a result, the JWST galaxies differ fundamentally from those in OU24: their redshift distribution extends to higher redshift on average, their SEDs contain a broader variety of continuum shapes and emission-line properties, and their selection functions are far from representative of a \textit{Roman} WL sample. This makes the JWST dataset unsuitable for direct shear-calibration forecasts but interesting for testing whether the regressions used to infer SED slopes behave consistently on real galaxies for different filter configurations.

Starting from the full NIRSpec compendium, we apply a series of quality cuts to construct a clean sample. We exclude stellar sources and require a signal-to-noise ratio $\mathrm{S/N} > 5$, consistent with previous spectroscopic continuum slope studies \citep[e.g.,][]{Saxena_2024}. The S/N here refers to the median S/N per spectral resolution element across the usable wavelength range of the extracted spectrum. In addition, we retain only spectra assigned the highest NIRSpec quality flag (grade = 3), which is reserved for objects with robust redshifts\footnote{See \url{https://dawn-cph.github.io/dja/blog/2025/05/01/nirspec-merged-table-v4/} for a detailed description of the NIRSpec grading scheme and examples of its application. We adopt the same grade selection as is used in the examples.}, and we remove duplicate observations by keeping the highest-S/N spectrum for each source ID. 
We further require the spectra to contain no missing data and restrict the sample to redshifts $z<3.06$ to match the maximum redshift in our OU24 sample. Photometry is obtained by integrating the spectra through LSST and \textit{Roman} filters; because NIRSpec has a minimum wavelength of $\sim 0.6~\mu$m, we compute synthetic LSST $r,i,z,y$ magnitudes but cannot obtain $u$ and $g$. This limitation is not critical for the chromatic PSF problem, since the dominant information about the NIR SED slope from optical filters resides in the redder optical bands. Finally, we apply an $H<25$ magnitude cut to ensure the sample resembles the WL selection we imposed on the OU24 galaxies we simulated. After all cuts, we retain 2{,}194 galaxies, and use an 80/20 train--test split to evaluate the RBF regression. Fig.~\ref{fig:jwst_cosmos_comp} shows the $H$-band magnitude and $J-H$ color evolution with redshift for our JWST NIRSpec sample compared to COSMOS2020 galaxies. We see that even though the galaxy sample is very different between both, the redshift evolution of the JWST NIR colors roughly matches that of COSMOS20. 

\begin{figure*}
    \centering

    
    \centering
    \includegraphics[width=\textwidth]{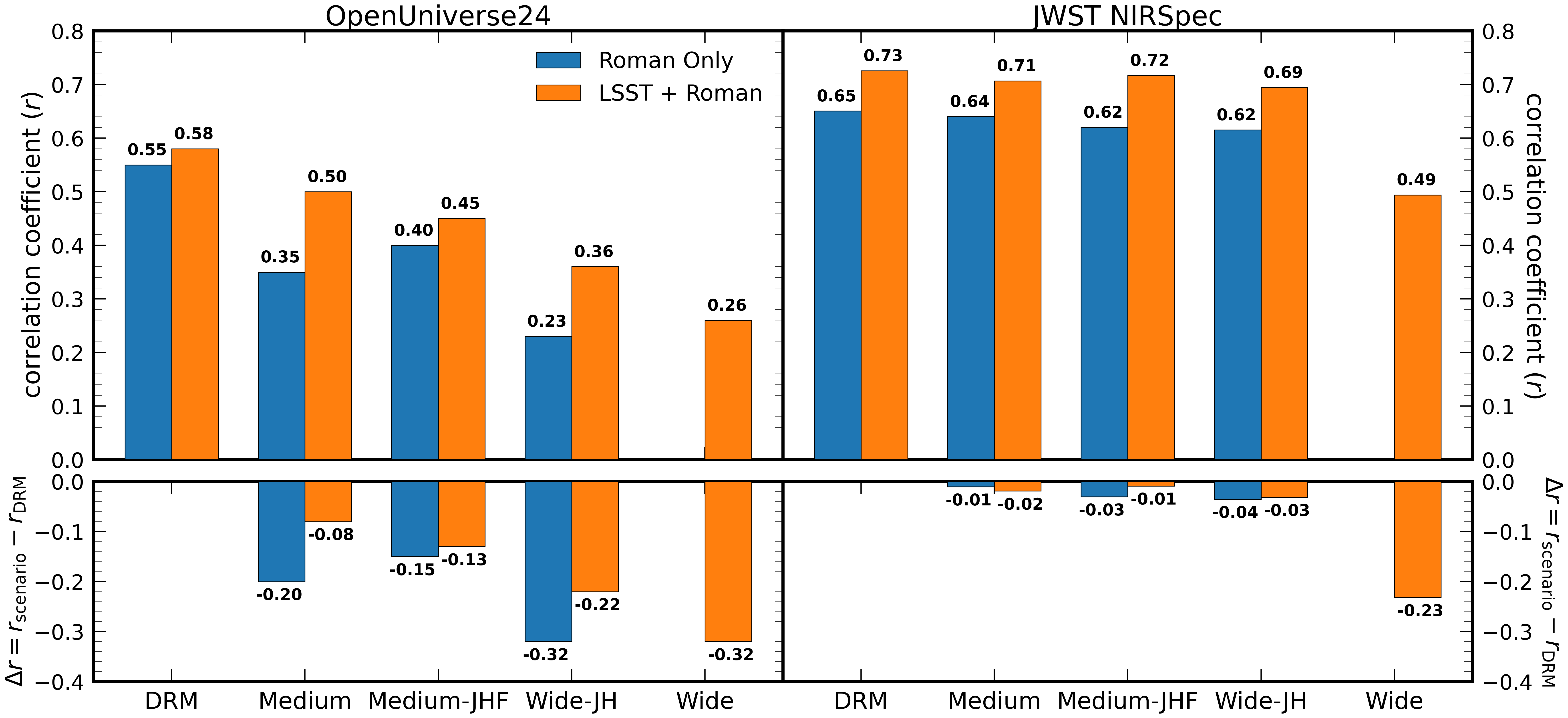}
    \caption{
    Performance of the RBF-based $H$-band SED slope inference across survey strategies, quantified by the Pearson correlation coefficient $r$ between predicted and true values (top panels). Results are shown for a representative OpenUniverse2024 training sample (left) and for a spectroscopic comparison sample from JWST NIRSpec (right). Blue and orange bars denote models trained and evaluated using LSST+\textit{Roman} photometry and \textit{Roman}-only photometry, respectively. The lower panels show the change in correlation relative to the DRM configuration, $\Delta r \equiv r_{\rm scenario}-r_{\rm DRM}$. Overall, some trends are consistent between OU24 and JWST, mostly that removing all NIR colors can drastically diminish the ability to infer the $H$-band SED slope, and that differences between either Medium tier configuration are small. JWST data seems to indicate smaller differences between scenarios with 2, 3, or 4 bands, with most of the information being carried by the $J-H$ color, while OU24 seems to exhibit a progressive degradation as NIR coverage is removed. Incorporating LSST photometry mildly improves performance in some scenarios for OU24, with the largest gains observed for the 2-band wide tier and 3-band medium tier, while the JWST dataset does not exhibit considerable improvements in performance when adding LSST photometry in either scenario.
}

    \label{fig:jwst_comp}

\end{figure*}

Figure~\ref{fig:jwst_comp} compares the correlation coefficients between the true and inferred $H$-band SED slopes across all survey strategies for both OU24 (left panel) and JWST (right panel). Because the JWST and OU24 galaxy populations differ strongly in redshift distribution, SED diversity, and selection function, we do not interpret absolute correlation values as performance metrics. Instead, we focus on relative trends across survey strategies and on the impact of including LSST photometry. The following conclusions summarize the main points of agreement and disagreement between the simulations and the JWST data. 

Overall, while OU24 predicts somewhat larger differences between near-infrared band combinations than are observed in the JWST spectroscopic sample, 
these differences do not translate into substantial changes in chromatic shear calibration performance. OU24 predicts a noticeable degradation in correlation when moving from four-band (YJHF) to three-band (YJH) and two-band (JH) NIR strategies, whereas the JWST sample shows much smaller differences between these configurations. This discrepancy can be traced to a strong empirical correlation in the JWST data between the $J-H$ color and the true $H$-band SED slope. For many JWST galaxies, the $J-H$ color alone captures most of the relevant spectral variation, rendering the additional Y-band information comparatively less informative. When combined with the shear-calibration results—which showed only modest differences between the DRM and medium-tier strategies—this supports the conclusion that the effective difference between four-band and three-band NIR surveys is likely small for chromatic correction.

Both OU24 and JWST indicate a modest benefit from including LSST photometry. In both datasets, LSST+\textit{Roman} photometry yields slightly higher correlations than \textit{Roman}-only photometry for most strategies, consistent with optical colors providing complementary information about the SED slope, particularly at higher redshift. While the medium-tier strategy in OU24 appears more sensitive to the inclusion of LSST than suggested by the JWST sample, the overall conclusion is consistent: \textit{Roman}-only regression remains viable for chromatic correction in the medium tier, with LSST offering incremental but non-essential improvements.

In contrast, both datasets agree on the significant loss of information when NIR color information is removed entirely, as in the wide tier (H-only) strategy. This configuration shows the steepest decline in correlation in both OU24 and JWST, despite their very different galaxy populations. This agreement reinforces the conclusion that the absence of NIR color information substantially limits the ability to recover the SED slope, and that a single-band weak-lensing survey can be under-constrained for chromatic correction.

Finally, OU24 and JWST differ in their assessment of the importance of LSST colors for a two-band (JH) survey. OU24 suggests that the JH strategy performs acceptably when LSST photometry is included but degrades significantly for \textit{Roman}-only data. In contrast, the JWST sample shows little to no degradation when LSST colors are removed. As with the three-band case, this difference reflects the tight $J-H$–SED-slope correlation present in the JWST data but absent in OU24. Consequently, while OU24 implies a more prominent role for LSST photometry in two-band surveys, the JWST results suggest that J and H alone may be sufficient for chromatic correction in real galaxies.

Overall, the JWST/NIRSpec comparison provides an important empirical cross-check on the simulation-based conclusions, though it is limited by the inhomogeneous nature of the NIRSpec dataset. The differences between the simulated and real datasets highlight the value of real-galaxy spectroscopy for validating color--SED relationships, while the points of agreement reinforce the main result of this work: chromatic correction for \textit{Roman} weak lensing is robust with three or more NIR bands, viable with two, and challenging in a single-band configuration.


\subsection{Summary of Shear Calibration Results}
The results from simulations, in combination with tests using a JWST spectroscopic sample, lead to the following conclusions:

\begin{itemize}[leftmargin=*]
    \item \textbf{Three- and four-band NIR strategies robustly mitigate chromatic shear calibration biases at the level of $|m| \lesssim 10^{-3}$.}  
    Minimal information is lost when removing the F184 filter, and no statistically significant differences are observed between the YJH and JHF three-band configurations. These results hold for both representative and bright training samples, indicating that three-band NIR coverage is sufficient for accurate chromatic correction. This conclusion is supported by our tests with JWST NIRSpec data.

    \item \textbf{Single-band (Wide-tier) strategies are intrinsically limited for chromatic correction, particularly with non-representative training samples.}  
    In the absence of NIR color information, residual shear biases can be larger and more sensitive to training-set mismatch. This highlights a fundamental limitation of single-band weak-lensing surveys for controlling chromatic PSF effects. The loss of information was similar for both the OU24 and JWST samples, thus bolstering the robustness of this conclusion.

    \item \textbf{An average color-based correction derived from the medium tier may provide a viable mitigation for the wide tier if estimated accurately.}  
    Applying an average correction may be sufficient, but can exceed requirements for individual tomographic bins. This approach provides a practical fallback in the absence of per-galaxy color information in the wide tier, and is most likely sufficient for other science cases than dark energy constraints with more relaxed shear bias requirements. 

    \item \textbf{Two-band (JH) Wide-tier strategies improve chromatic correction relative to a single-band configuration, but the extent is unclear.}  
    The addition of a second NIR band reduces residual shear biases and improves robustness to training-sample mismatch. However, performance remains inferior to three-band strategies, and whether this improvement justifies the corresponding loss in survey area depends on the adopted systematic error budget. Our tests with JWST NIRSpec data suggests the the majority of the information is contained in the $J-H$ color, indicating a single color might be good enough to correct for chromatic effects.

    \item \textbf{LSST photometry provides performance improvements, but is not essential for medium-tier chromatic correction.}  
    Optical photometry generally tightens residual biases, however, for three- and four-band NIR configurations, \textit{Roman}-only photometry might be sufficient to meet shear-calibration requirements. The fact that biases exceed relaxed requirements for at most one tomographic bin indicates that a \textit{Roman}-only correction for the medium tier is viable. This conclusion is supported by our tests with JWST NIRSpec data, where we did not observe a notable difference when adding LSST photometry.
\end{itemize}

Taken together, these results indicate that the ability to meet \textit{Roman}’s WL systematics budget depends sensitively on the availability of NIR color information and the representativeness of the training sample. Multiband NIR imaging provides a clear path to robust chromatic mitigation, while single-band surveys present significant challenges that may be difficult to overcome without additional calibration resources.

\section{Cosmological Inference}\label{Cosmology}

The ultimate requirement for any chromatic PSF mitigation scheme is not merely that it reduce per-bin shear biases, but that it enable unbiased cosmological inference at the precision targeted by \textit{Roman}. In this section, we propagate the residual multiplicative shear biases measured in Section~\ref{Shear_Cal} into a cosmological analysis, quantifying their impact on constraints on $S_8$. Our goal is to assess whether different survey strategies and correction schemes recover unbiased cosmological parameters, and to identify configurations for which chromatic effects remain a limiting systematic. In this section we only analyze the results of a \textit{Roman}-only correction, since the results from shear calibration seemed to indicate the biases were mostly within requirements except for a few cases. Given the computational simplicity of a \textit{Roman}-only correction and the performance demonstrated above, we are interested in assessing the impact of the residual biases in cosmological inference.

\subsection{Inference framework and setup}

We perform cosmological inference assuming a spatially flat $\Lambda$CDM model, using the Cobaya--CosmoLike Joint Architecture\footnote{\url{https://github.com/CosmoLike/cocoa}} (\textsc{Cocoa}; Miranda et al.\ 2026, in preparation), 
which enables likelihood-based parameter estimation from weak-lensing observables while consistently incorporating cosmological, astrophysical, and observational nuisance parameters, including shear calibration and photometric redshift uncertainties. We also use a neural network emulator trained on CoCoA model vectors to boost the MCMC chains. The architecture of the neural network and training methods are developed in \cite{Zhong_2025, Saraivanov_2025, Xu_2025}, and details specific to this emulator are presented in Xu et al. (in prep). 

Our analysis builds on the data vectors, masks, covariance matrices, and redshift distributions ($n(z)$) generated by the Cosmological Parameters Inference Pipeline (CPIP) working group in the \textit{Roman} HLIS Cosmology Project Infrastructure Team (PIT) as part of their first internal Data Challenge (DC1).  The DC1 products correspond to the \textit{Roman} medium-tier survey configuration and include the three two-point statistics between galaxy clustering, weak lensing, and galaxy-galaxy lensing (so-called $3\times2$-pt) in both real and Fourier space. DC1 adopts theoretical mock data vectors calculated at blinded fiducial model parameters and utilizes an analytical non-Gaussian covariance matrix calculated by \textsc{CosmoCov} \citep{Krause_2017, Fang_2020}. 

We adopt the Fourier space tomographic angular power spectra to explore the bias in $\Omega_{\mathrm{m}}$--$S_8$ in this work. The angular power spectra include 15 logarithmic bins between $l=30$ and $l=4000$. We choose an aggressive scale cut to include all data points, except for galaxy-galaxy lensing, where we discard the last angular bin ($l>2886$) for a positive-definite covariance matrix. This is aimed at stress testing the impact of shear calibration bias due to chromatic PSF effect, and deriving a robust scale cut is beyond the scope of this work. In addition, DC1 assumes that the same galaxy sample is used for both shear and galaxy position measurements. The full parameter space includes five cosmological parameters $(\Omega_{\mathrm{m}}, A_s, n_s, \Omega_{\mathrm{b}}, h_0)$, along with nuisance parameters describing galaxy bias, intrinsic alignments, photometric redshift biases, and multiplicative shear calibration biases. All parameters are varied and marginalized over in the inference.

\begin{table}
\footnotesize
\centering
\caption{Baseline inference setup for the $\Lambda$CDM \textit{Roman} medium tier Fourier-space analysis. Listed are fiducial parameter values and priors. Uniform and Gaussian priors are represented by $U[\min,\max]$ and $\mathcal{N}(\mu,\,\sigma)$, respectively.}
\label{tab:cosmo_setup}
\begin{tabular}{l l l}
\hline\hline
Parameter & Prior & Fiducial value \\
\hline
\multicolumn{3}{l}{\textbf{Cosmology ($\Lambda$CDM)}} \\
$\Omega_{\mathrm{m}}$  & $U[0.1,\,0.9]$ & 0.3156 \\ 
$A_s \times 10^9$      & $U[0.5,\,5]$   & 2.10 \\
$\sigma_8$            & Derived        & 0.8255 \\ 
$n_s$                 & $U[0.87,\,1.07]$ & 0.9645 \\
$\Omega_{\mathrm{b}}$            & $U[0.03,\,0.07]$ & 0.0492 \\
$h_0$                 & $U[0.55,\,0.91]$ & 0.6727 \\
\hline
\multicolumn{3}{l}{\textbf{Galaxy bias}} \\
$b^1$  & $U[0.8,\,3.0]$ & 1.18 \\
$b^2$  & $U[0.8,\,3.0]$ & 1.40 \\
$b^3$  & $U[0.8,\,3.0]$ & 1.55 \\
$b^4$  & $U[0.8,\,3.0]$ & 1.71 \\
$b^5$  & $U[0.8,\,3.0]$ & 1.90 \\
$b^6$  & $U[0.8,\,3.0]$ & 2.15 \\
$b^7$  & $U[0.8,\,3.0]$ & 2.52 \\
$b^8$  & $U[0.8,\,4.0]$ & 3.44 \\
\hline
\multicolumn{3}{l}{\textbf{Photometric redshift systematics}} \\
$\Delta z_i$ (all bins) & $\mathcal{N}(0,\,0.002)$ & 0.0 \\
\hline
\multicolumn{3}{l}{\textbf{Shear calibration systematics}} \\
$m_i$ (all bins) & $\mathcal{N}(0,\,0.005)$ & 0.0 \\
\hline
\multicolumn{3}{l}{\textbf{Intrinsic alignments (NLA)}} \\
$A_{\rm IA}$     & $U[-5,\,5]$ & 0.6061 \\
$\eta_{\rm IA}$  & $U[-5,\,5]$ & $-1.515$ \\
\hline\hline
\end{tabular}
\end{table}

We note that the wide tier described in the \citetalias{ROTAC_2025} is expected to be significantly shallower than the medium tier, with an effective number density of $n_{\mathrm{eff}} = 26.7~\mathrm{arcmin}^{-2}$ compared to $41.3~\mathrm{arcmin}^{-2}$, and to cover a substantially larger area. 
A realistic cosmological analysis of the wide tier would therefore require a modified covariance matrix and redshift distribution from that of the DC1 medium tier. Incorporating these changes is beyond the scope of this work, so we do not include the wide-tier in the cosmological inference presented here. Instead, we focus on propagating the biases presented in Sec.~\ref{Shear_Cal} into parameter constraints for the medium tier, and leave a dedicated cosmological analysis of the wide tier for future work. 

\begin{figure*}
    \centering

    \begin{subfigure}[b]{0.48\textwidth}
        \centering
        \includegraphics[width=\textwidth]{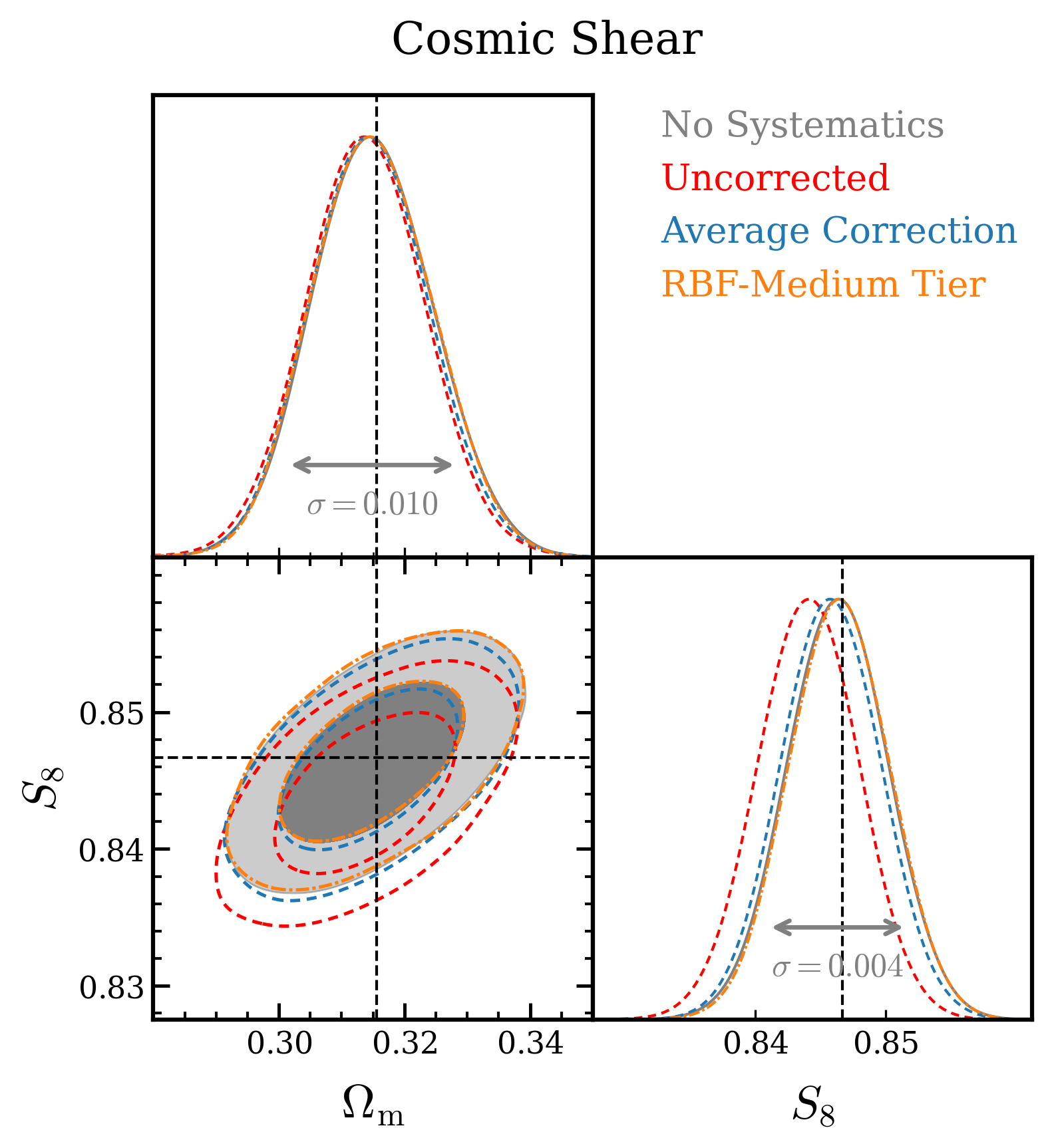}
    \end{subfigure}
    \begin{subfigure}[b]{0.48\textwidth}
        \centering
        \includegraphics[width=\textwidth]{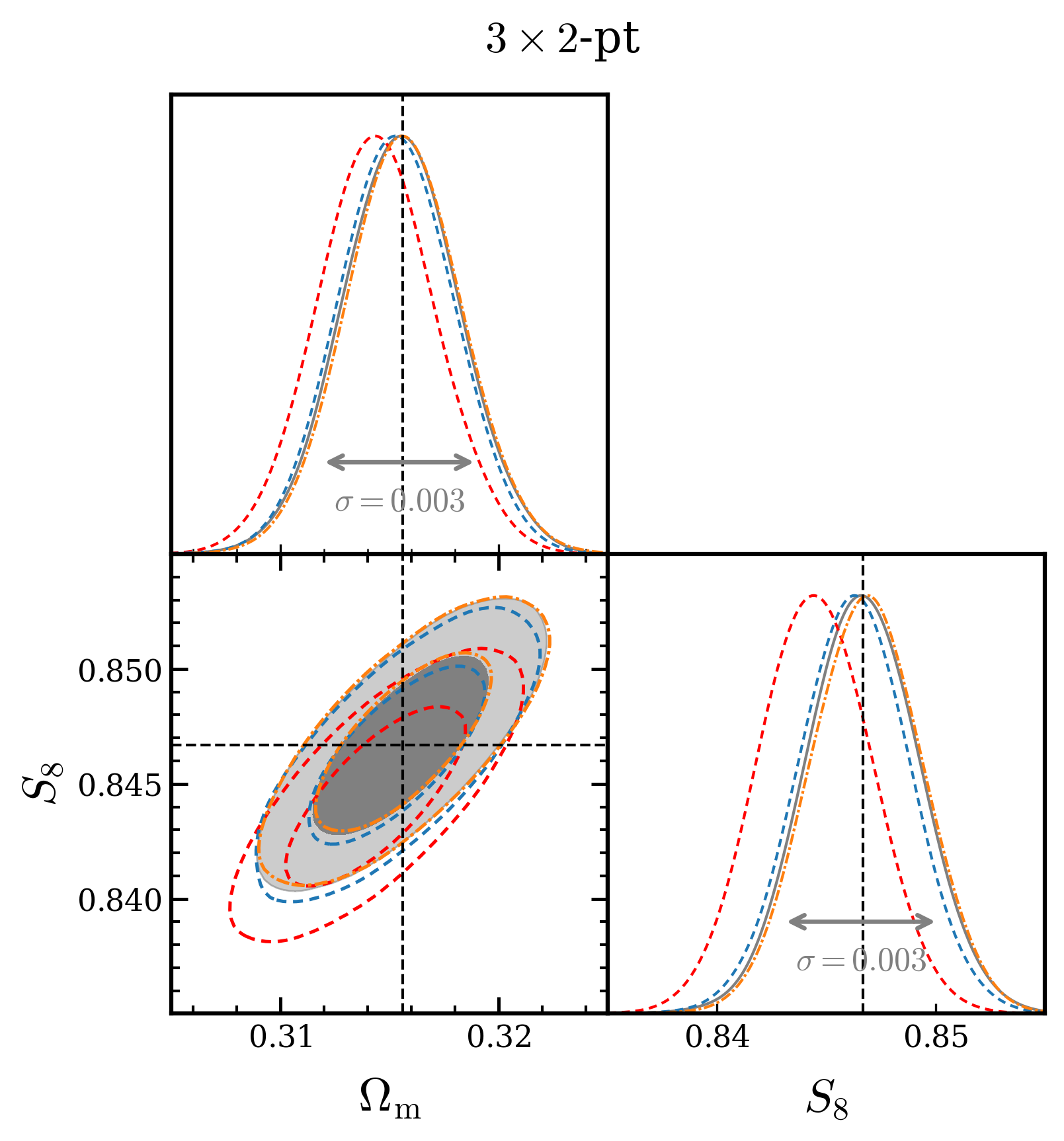}
    \end{subfigure}
    \caption{
    Posterior constraints in the $\Omega_{\mathrm{m}}$–$S_8$ plane for cosmic shear (left) and joint $3\times2$-point analyses (right) in a $\Lambda$CDM analysis, illustrating the impact of shear calibration uncertainty due to chromatic PSF systematics under different correction schemes for a full survey dataset. Gray contours indicate the no-systematics reference case, while red dashed contours show the case of uncorrected chromatic shear bias. Blue contours correspond to applying a single average correction, and orange contours show RBF-based corrections (from a bright training sample) derived from survey-specific SED slope inference for the medium tier. 
    One- and two-dimensional marginalized posteriors are shown at the 68\% and 95\% confidence levels. RBF-based corrections substantially reduce the bias in both $\Omega_{\mathrm{m}}$ and $S_8$ relative to the uncorrected case, recovering constraints close to the no-systematics baseline.} 

    \label{fig:corner_plots}

\end{figure*}

\begin{table*}
    \centering
    \caption{
    Bias in the derived $S_8$ for different chromatic mitigation schemes and survey strategies.
    Biases are shown for cosmic shear-only and 3$\times 2$-pt analyses, in units of
    $\sigma(S_8)$, the standard deviation of $S_8$ in the reference case without chromatic systematics. The top block lists results for the uncorrected
    and average-SED correction cases.
    The bottom block lists results for the RBF-based correction using either a
    representative or bright training sample for each survey strategy. The priors on the multiplicative bias are $\sigma_m = 0.005$.
    }
    \label{tab:s8_bias}
    \begin{tabular}{lcccc}
        \hline
        & \multicolumn{2}{c}{Cosmic shear only}
        & \multicolumn{2}{c}{3$\times 2$-pt} \\
        \cline{2-3} \cline{4-5}
        Scheme / Strategy
        & Rep.\ sample & Bright sample
        & Rep.\ sample & Bright sample \\
        \hline
        Uncorrected
        & \multicolumn{2}{c}{$-0.58 \sigma$}
        & \multicolumn{2}{c}{$-0.70 \sigma$} \\
        Average SED correction
        & \multicolumn{2}{c}{$-0.13 \sigma$}
        & \multicolumn{2}{c}{$-0.14 \sigma$} \\
        \hline
        DRM
        & $0.02 \sigma$ & $0.04 \sigma$
        & $0.05 \sigma$ & $0.04 \sigma$ \\
        Medium
        & $-0.04 \sigma$ & $-0.01 \sigma$
        & $-0.07 \sigma$ & $-0.01 \sigma$ \\
        Medium--JHF
        & $0.01 \sigma$ & $0.06 \sigma$
        & $0.06 \sigma$ & $0.15 \sigma$ \\
        \hline
    \end{tabular}
\end{table*}

To isolate the impact of residual multiplicative shear biases, we apply two modifications to the original DC1 data vector to define a new baseline for this study. First, we set the multiplicative shear bias to zero in all tomographic bins. Second, we similarly set all photometric redshift bias parameters to zero. These changes are applied because the original DC1 data vector includes non-zero values for these nuisance parameters; redefining the baseline in this way allows us to cleanly introduce controlled shear calibration offsets while keeping the underlying cosmology, galaxy bias, and intrinsic alignment model fixed. We adopt the same priors on all cosmological and nuisance parameters as those used in DC1. The fiducial values and priors for all cosmological and nuisance parameters are summarized in Table~\ref{tab:cosmo_setup}. We note that the prior width on shear biases, $\sigma_m = 0.005$, differs from the relaxed requirements discussed in Sec.~\ref{BiasRequirements}. The prior width and shear bias requirements are not defined or set equivalently, with the prior width being related to how well we think we can use simulations and other means to constrain the systematic uncertainties.  Deriving a well-motivated and realistic prior on multiplicative shear bias for \textit{Roman} is out of the scope of this work. Therefore, we adopt the DC1 priors and defer this question to future work. 

We analyze two data-vector configurations: a cosmic-shear-only analysis and a joint $3\times2$-point analysis combining cosmic shear, galaxy--galaxy lensing, and galaxy clustering. These represent two key analyses planned for the \textit{Roman} HLIS 
and allow us to assess how chromatic systematics propagate in analyses of differing statistical power and power to self-calibrate systematics. We marginalize over the whole parameter space and focus on the $\Omega_{\mathrm{m}}$--$S_8$ plane, where
\begin{equation}
S_8 \equiv \sigma_8 \left( \frac{\Omega_{\mathrm{m}}}{0.3} \right)^{0.5}.
\end{equation}
This parameter combination corresponds to the best-constrained direction in the degenerate $\Omega_{\mathrm{m}}$–$\sigma_8$ parameter space probed by weak-lensing measurements, and is therefore particularly sensitive to multiplicative shear biases. 

\subsection{Cosmological impact of chromatic mitigation}

We propagate the residual multiplicative shear biases measured in Section~\ref{Shear_Cal} into a full cosmological inference analysis for the medium-tier survey strategies and mitigation schemes by contaminating the baseline data vectors with the measured shear biases. We run a full Markov Chain Monte Carlo (MCMC) analysis for every configuration listed in Table~\ref{tab:s8_bias}, including both cosmic-shear-only and joint $3\times2$-point analyses. The MCMC is considered to have converged when the generalized Gelman-Rubin statistic \citep{Gelman_1992}, $R-1$, reaches a value of $0.02$ \citep{Lewis_2013}. For clarity, we present posterior contours only for the bright training sample of the ROTAC-recommended medium tier, the uncorrected case, and the average-correction scheme, while summarizing the resulting $S_8$ biases in Table~\ref{tab:s8_bias}. We show \textit{Roman}-only corrections as the residual biases shown in Sec.~\ref{Shear_Cal} remain approximately within requirements, and because, as discussed in Sec.~\ref{sec:lsst_photometry}, incorporating LSST photometry at early stages of the image-processing pipeline is non-trivial. Consequently, a \textit{Roman}-only approach is preferable if it is sufficient to achieve unbiased cosmological constraints. 

Figure~\ref{fig:corner_plots} illustrates the resulting posterior constraints in the $\Omega_{\mathrm{m}}$–$S_8$ plane for cosmic shear (left) and joint $3\times2$-point analyses (right). In the absence of any chromatic correction, the inferred cosmological parameters are significantly biased relative to the no-systematics baseline, with shifts of $|\Delta S_8| \gtrsim 0.5\sigma$ for cosmic shear and $0.7\sigma$ for the $3\times2$-point analysis. This demonstrates that uncorrected chromatic PSF effects would represent a limiting systematic for \textit{Roman} weak lensing cosmology, even when relatively broad priors on the multiplicative bias are adopted.

Applying a single average correction substantially reduces these biases. In both cosmic shear and $3\times2$-point analyses, the average correction recovers constraints within $\sim 0.1\sigma$ of the no-systematics case. While this level of residual bias may be acceptable depending on the final systematic error budget, it is not fully eliminated. For the medium tier configuration, both cosmic shear and $3\times2$-point analyses show negligible residual bias, consistent with the shear-calibration results in Section~\ref{Shear_Cal}. 

Table~\ref{tab:s8_bias} provides a quantitative summary of these effects across all other survey strategies and correction schemes. Both the DRM and medium tier strategies reduce the bias in $S_8$ to $\lesssim 0.07\sigma$ for both cosmic shear and $3\times2$-point analyses (with the exception of the $3\times2$pt analysis of the bright sample with the JHF medium configuration), even when the regression is trained on a bright, non-representative sample. Once again, we see small differences between the medium tier (either with YJH or JHF) and the DRM, as well as no systematic preference between either of the 3-band medium tiers.

We note that discrepancies between the inferred and true cosmological parameters in the presence of non-zero multiplicative shear biases arise from the mismatch between the injected non-zero bias values and zero-centered priors on those parameters. In the absence of noise, the contaminated data vector can be reproduced by evaluating the model at the true cosmological and nuisance parameters, yielding a maximum likelihood at that point. The differences between the posterior mean values and the true parameters therefore show the influence of the priors center on the nuisance parameters rather than a failure of the likelihood to describe the contaminated data. This interpretation is consistent with the findings of \citet{Cao_2026}.


In summary, residual chromatic PSF effects can propagate into non-negligible biases in cosmological parameters if left uncorrected, with shifts in $S_8$ of $\sim 0.5-0.7\sigma$ in cosmic shear and $3\times2$-point analyses,  in a $\Lambda$CDM cosmology analysis that marginalizes over many sources of systematic uncertainty. Simple average corrections substantially reduce these biases but typically leave residual offsets at the $\sim 0.1\sigma$ level. The per-galaxy RBF-based correction for the medium tier survey recovers cosmological constraints with residual biases $\lesssim 0.07\sigma$. Single-band configurations remain the most challenging, exhibiting larger residual biases that could require further calibration.

\section{Conclusion}\label{Conclusion}

In this work, we have investigated how survey design choices for the \textit{Roman} High Latitude Wide Area Survey affect the mitigation of chromatic PSF effects in weak lensing shear measurements. Building on the PSF-level mitigation framework developed in \citetalias{Berlfein_2025}, we examined how the availability of NIR color information, the representativeness of the training sample, and the inclusion of auxiliary optical photometry influence our ability to infer galaxy SED slopes and correct wavelength-dependent PSF biases. Using realistic image simulations, regression-based inference, validation against JWST NIRSpec data, and full cosmological inference, we quantified the implications of these choices for both shear calibration and downstream cosmological constraints.

The primary questions posed in Section~\ref{sec:intro} can be answered as follows:

\begin{itemize}[leftmargin=*]
    \item \textbf{4 vs.\ 3 bands: Does the removal of the F184 filter in the medium tier substantially affect our ability to correct for chromatic PSF effects?}  
    No. We find minimal loss of performance when moving from the four-band DRM (YJHF) to three-band medium tier configurations. Residual shear biases remain within relaxed requirements (Sec.~\ref{BiasRequirements}), and cosmological constraints are effectively unbiased. This conclusion is supported by both simulation-based results and empirical validation using JWST NIRSpec spectra.

    \item \textbf{Which 3 bands: In the case of a 3-band medium tier, is there an important difference between YJH and JHF?}  
    No statistically significant difference is observed. Both configurations provide sufficient NIR color information to infer the H-band SED slope accurately, leading to comparable shear calibration performance and cosmological impact. This indicates that the specific choice of the third NIR band is not critical for chromatic mitigation, provided that three-band coverage is available.

    \item \textbf{Single band correction: How well can we correct for these effects for a single-band wide tier, either by using LSST photometry or by applying an average correction derived using the medium tier?}  
    Single-band (H-only) strategies are intrinsically limited for chromatic correction. While RBF-based per-galaxy corrections derived using LSST photometry reduce biases substantially relative to the uncorrected case, residual shear and cosmological biases persist, particularly for non-representative training samples. These limitations reflect the fundamental loss of SED information when NIR color constraints are absent. 
    
    We also achieved partial success in deriving an average color-based correction from the medium tier and applying it to the wide tier. This reduced shear biases within requirements for most tomographic bins, and in a medium tier cosmological analysis to the $\sim 0.1\sigma$ level in $S_8$. Its effectiveness in a wide tier cosmological forecast is left for future work.

    \item \textbf{Is two bands enough: Is a 2-band wide tier (JH) enough to correct for these effects?}  
    A two-band wide tier configuration performs significantly better than a single-band survey, reducing residual shear biases by roughly a factor of two and approaching the performance of three-band strategies. However, it remains less robust than full medium tier coverage, and whether this improvement justifies the corresponding reduction in survey area depends on the adopted systematic error budget. Validation with JWST data suggests that much of the relevant information is carried by the $J-H$ color, indicating that a single NIR color derived from two passbands can already provide substantial leverage.

    \item \textbf{Need for LSST: Will we need  LSST photometry to apply an accurate correction in the medium and/or wide tier?}  
    LSST photometry provides modest but consistent improvements across most configurations, particularly for the wide tier and two-band strategies. However, for three- and four-band NIR surveys, \textit{Roman}-only photometry is generally sufficient to achieve accurate chromatic correction. Both simulations and JWST validation indicate that LSST photometry is beneficial but not essential for medium tier chromatic mitigation.

     It is important to emphasize that this conclusion applies specifically to chromatic PSF correction and does not address the separate question of photometric redshift estimation. In practice, LSST photometry is still likely to play an important role in photo-$z$ inference, where incorporating external optical data occurs at a later stage of the analysis pipeline and is therefore operationally simpler than integrating it directly into the PSF modeling stage.
\end{itemize}

Taken together, these results demonstrate that the ability to control chromatic PSF systematics for \textit{Roman} weak lensing depends primarily on the availability of NIR color information and the representativeness of the training data used for SED inference. Multiband NIR imaging, particularly three-band coverage, provides a robust and flexible path to meeting shear calibration and cosmological requirements, even under realistic training conditions. In contrast, single-band strategies face intrinsic limitations that are difficult to overcome without additional calibration assumptions or auxiliary data.

Looking forward, several extensions of this work will be important. Incorporating fully realistic image simulations, including noise, blending, detection effects, and coadded image products, will be necessary to assess how chromatic mitigation performs in end-to-end \textit{Roman} analyses. Further exploration of hybrid strategies, such as using the medium tier corrections to inform a  wide tier correction will help refine survey optimization trade-offs. Finally, continued validation against real galaxy spectroscopy will be essential for comparing simulation-based forecasts to real galaxy populations.

Finally, it is important to emphasize that many aspects of chromatic PSF mitigation will ultimately be informed by real \textit{Roman} data. The true galaxy population, observing conditions, and instrumental behavior will inevitably introduce complexities that cannot be fully captured by simulations alone. In addition, the extent to which the \textit{Roman} High Latitude Spectroscopic Survey can contribute to calibrating these effects remains an open question. Nevertheless, the framework developed here provide a robust foundation for interpreting these effects once \textit{Roman} is operational. By identifying which survey design choices most strongly impact chromatic systematics, and by quantifying their cosmological consequences, this work helps clarify the key questions that will need to be addressed with early \textit{Roman} data and establishes the tools required to explore them more rigorously.


\section*{Acknowledgments}

This paper has undergone internal review in the \textit{Roman} High Latitude Imaging Survey Cosmology Project Infrastructure Team (PIT). We would like to thank Kaili Cao and Mike Jarvis for helpful comments and feedback during the review process.

This work was supported in part by the OpenUniverse effort, which is funded by NASA under JPL Contract Task 70-711320, “Maximizing Science Exploitation of Simulated Cosmological Survey Data Across Surveys”; and in part by the “Maximizing Cosmological Science with the \textit{Roman} High Latitude Imaging Survey” \textit{Roman} Project Infrastructure Team (NASA grant 22-ROMAN11-0011). JX is also supported by the \textit{Roman} Research and Support Participation program (NASA grant 80NSSC24K0088). TZ is supported by Schmidt Sciences.

\section*{Data Availability}
The \texttt{Diffsky} extragalactic catalog and stellar catalog are available through the NASA/IPAC
Infrared Science Archive (IRSA) at \url{https://irsa.ipac.caltech.edu/data/theory/openuniverse2024/overview.html}. The \texttt{GalSim} package is publicly available at \url{https://github.com/GalSim-developers/GalSim}. The \texttt{PhotErr} package is publicly available at \url{https://github.com/jfcrenshaw/photerr}. The code used to generate the image simulations and analysis for this work is available at \url{https://github.com/Roman-HLIS-Cosmology-PIT/RomanChromaticPSF}.



\bibliographystyle{mnras}
\bibliography{main}





\bsp	
\label{lastpage}
\end{document}